\pdfoutput=1
\documentclass{aa}  
\def\eCBSMBH{{\it e}CB-SMBH}
\def\Rsub{R_{\rm sub}}

\usepackage{subcaption}

\usepackage{graphicx}
\usepackage{txfonts}
\usepackage{hyperref}
\begin{document}

   \title{Detection of eccentric close-binary supermassive black holes with incomplete interferometric data}

\author{Andjelka B. Kova{\v c}evi{\'c}  \inst{1,2},
        Yu-Yang Songsheng  \inst{3},
        Jian-Min Wang \inst{3,4}, 
        Luka  {\v C}. Popovi{\'c}  \inst{1,5,2}  
}

\institute{Department of astronomy, Faculty of mathematics, University of Belgrade
        Studentski trg 16, Belgrade, 11000, Serbia\\
        \email{andjelka@matf.bg.ac.rs}
        \and 
        PIFI Research Fellow, Key Laboratory for Particle Astrophysics, Institute of High Energy Physics, Chinese Academy of Sciences,19B Yuquan Road, 100049 Beijing, China
        \and
        Key Laboratory for Particle Astrophysics, Institute of High Energy Physics, CAS
        19B Yuquan Road, Beijing 100049, China\\
        \and
        {School of Astronomy and Space Sciences, University of Chinese Academy of Sciences, Beijing 100049, China}\\
        \email{wangjm@ihep.ac.cn}
        \and
        Astronomical observatory Belgrade 
        Volgina 7, P.O.Box 74 11060, Belgrade,  11060, Serbia\\
}
   \date{Received; accepted}

 
  \abstract
   {Recent studies have proposed that {General Relativity Analysis via VLT InTerferometrY}  upgrades  (GRAVITY+) on board the
{Very Large Telescope Interferometer} (VLTI) are able to trace the circular orbit of the subparsec ($\lesssim 0.1$ pc) close-binary supermassive black holes (CB-SMBHs) by measuring the photo-centre variation of the hot dust emission. However, the CB-SMBHs orbit may become highly eccentric throughout the evolution of these objects, and the orbital period may be far longer than the observational time baseline.}
   {We investigate  the problem of detecting the CB-SMBH with hot dust emission and high eccentricity ({\eCBSMBH}, $e=0.5$) when the observed time baselines of their astrometric data and radial velocities are considerably shorter than the orbital period.}
   {The parameter space of the Keplerian model of the {\eCBSMBH} is large for exploratory purposes. We therefore applied the Bayesian method to fit orbital elements of the {\eCBSMBH} to  combined radial velocity and astrometric data covering a small fraction of the orbital period.}
  { We estimate that a number of potential {\eCBSMBH} systems within reach of GRAVITY + 
  will be {similar to the number of}  
  planned circular targets.
        We show that using observational time baselines that cover  $\gtrsim 10\%$ of the orbit increases the possibility of determining the period, eccentricity, and total mass of an \eCBSMBH.  When the observational time baseline becomes too short ($\sim 5\% $), the quality of the retrieved\, {\eCBSMBH}\, parameters degrades.
We also illustrate how interferometry may be used to estimate the photo-centre at the {\eCBSMBH} emission line, which could be relevant for GRAVITY+ successors.
    Even if the astrometric signal for \eCBSMBH\, systems is reduced by a factor of $\sqrt{1-e^{2}}$  compared to circular ones,  we find that the hot dust emission of {\eCBSMBH}s can be traced by GRAVITY+ at the elementary level.}
    {}
  
 \titlerunning{{\eCBSMBH} Interferometry}
   \authorrunning{Kova{\v c}evi{\'c} et al.}
   \maketitle
  
%

\section{Introduction}\label{sec:intro}

 It is now well known that almost all galaxies contain supermassive black holes (SMBHs) at their cores \citep{10.1088/0004-637X/695/2/1577,Kormendy2013}, with SMBH masses in the range $10^{5}- 10^{9.5} M_{\odot}$ \citep{10.1111/j.1365-2966.2012.21651.x}.
 Mergers of galaxies unavoidably lead to the formation of SMBH binaries  
 \citep[SMBHBs;][]{10.1038/287307a0, 10.1086/323830}. As galaxy mergers have been shown to funnel considerable amounts of gas to the nuclear area \citep{10.1086/504412}, binaries are expected to be surrounded by gas. This phenomena spurred a quest for detecting SMBHBs that may accrete gas and release variable bright electromagnetic emission due to their dynamic interplay with the surrounding gas \citep[see review by][]{Bogd21}{;} and various imprints of 
 electromagnetic signatures  of dual  and  binary SMBH candidates were found, 
 with separations from $\sim$ 1 kpc to subparsec values
 \citep[see exhaustive reviews of ][]{10.1016/j.newar.2011.11.001, 10.1088/0004-637X/785/2/115, 10.1016/j.newar.2020.101525, Bogd21}.
 {
        Although dozens of dual {active galactic nuclei} (AGNs) have been spatially resolved, subparsec SMBHBs
have remained elusive due to controversial electromagnetic characteristics \citep{2022MNRAS.510.5929C}.
        In addition, particular effort has been made to obtain observational evidence for SMBHBs with subparsec separations of 0.1 pc, known as close-binary  SMBHs \citep[CB-SMBHs,][]{10.1088/1674-4527/20/10/160}, built on the notion that they are viable nanohertz (nano-Hz) gravitational wave (GW) sources.
        Once the binary has reached subparsec scales, the SMBHs can spiral together and combine over timescales of less than the age of the Universe \citep{10.1038/287307a0}. SMBHBs become significant gravitational wave generators in the final months or years before merger, and could be detected by pulsar timing arrays \citep{10.1088/0264-9381/27/8/084013,  10.1093/mnras/stz2857}. The discovery of these binaries and the measurement of their orbital parameters would, without a doubt, be extremely beneficial in our efforts to detect nano-Hz gravitational waves in the nearby future \citep{10.1088/0264-9381/26/9/094033}.}
 {
        To understand the ultimate destiny  of an SMBH binary,  not only the orbital decay but also the eccentricity evolution of the pair must be investigated \citep{10.1155/2012/940568}. For a CB-SMBH with eccentricity ($e$: hereafter {\eCBSMBH}), the decay (or inspiral) timescale driven by only GW emission is given by}
 \begin{equation}\label{Eq:tgw}
 t_{\rm gw}=\frac{5}{256}\frac{c^{5}{a}^{4}}{G^{3}\mu M^{3}}F(e)^{-1}
 \approx 6\times 10^{10}a^{4}_{0.1}\mu^{-1}M_8^{-3} F(e)^{-1}\,{\rm yr}
 ,\end{equation}
 
 \noindent\citep{P64}, where $G$ is the gravitational constant, $c$ is the speed of light, $M=M_{1}+M_{2}$ where $M_1$ and $M_2$ are the masses of the primary and secondary SMBHs, 
 $\mu=M_{1}M_{2}/M^2$, ${M_{8}}=M/10^{8}M_{\odot}$,
 $a_{0.1}=a/{0.1\,{\rm pc}}$ is the semi-major axis in units of 0.1\,pc, $e$ is the orbital eccentricity, and 
 \begin{equation}\label{Eq:Fe}
 F(e)=\left(1+\frac{73}{24}e^{2}+ \frac{37}{96}e^{4}\right)\left(1-e^{2}\right)^{-7/2}   
 \end{equation}
 is enhancement factor which  increases with eccentricity.   
 Because $F(e)$ increases monotonically as eccentricity increases, we find from Eqs.(\ref{Eq:tgw}) and (\ref{Eq:Fe}) that the inspiral timescale of a binary can be shorter than that of the circular case (i.e. $F(0)^{-1}=1$). Also,  the inspiral time\footnote{On that timescale,  the eccentricity reduces as well \citep{Bogd21}. Because the velocity of components is greater at the pericentre, binaries emit more gravitational waves while in pericentre than when in apocentre. Due to this asymmetric emission of gravitational radiation, the orbit of a binary changes from elliptical to circular  \citep{Bogd21}. } is proportional to $\sim a^{4}_{0.1}$. 
 Because the mutual separations and eccentricities of {\eCBSMBH}s affect the inspiral time, {\eCBSMBH}s have recently become crucial for a wide range of studies, from black hole formation to gravitational wave physics \citep{10.3847/1538-4357/abad31}.

 Subparsec binary separations are typical in late-stage galactic mergers where two SMBHs are close enough to form a gravitationally bound system.
 The key theoretical  feature of CB-SMBHs is that their electromagnetic signatures could be related to the orbital elements of their motion  \citep{10.1051/0004-6361:20040314}, but they are observationally elusive due to their small separation on the sky, as well as the uncertainties related to the uniqueness of their observational signatures. 
 In addition, they are  expected to be inherently scarce, as their  occurrence   relies on their unknown evolutionary rate on small scales;  it is possible  {that a fraction } $(<0.001)$ {of}  AGNs at redshift $z<0.7$ may harbour  CB-SMBH 
 \citep{10.1088/0004-637X/703/1/L86, 10.1016/j.newar.2020.101525}. 
 Consequently, any observational search for CB-SMBHs must include a large sample of their host active galaxies and  must discriminate signatures of binaries from those  AGNs powered by a single SMBH.
 
 So far, observational searches for such systems have primarily focused on photometry and spectroscopic data, and rarely on direct imaging \citep[see e.g.][]{10.1016/j.newar.2020.101525}.
 For example, if CB-SMBHs hosted by active galaxies are made up of two distinct broad-line regions 
 \citep[BLRs; see e.g.][]{2000SerAJ.162....1P,10.1093/mnras/stab1510,  10.1088/0004-637X/725/1/249}, they might be studied using either reverberation mapping (RM) of their nuclear region  \citep{10.3847/1538-4357/aacdfa,10.1051/0004-6361/201936398,2020ApJS..247....3S}, or a long-term monitoring campaign of profile variations \citep[e.g.][]{
        10.1088/0067-0049/201/2/23,
        10.1088/0004-637X/777/1/44,
        10.3847/0004-637X/822/1/4,
        10.3847/0004-637x/817/1/42,
        10.3847/1538-4357/aaeff0,
        10.3847/1538-4357/ab88b5,
        10.3847/2041-8213/abb2ab,
        10.1088/0004-637x/789/2/140,
        10.1088/0067-0049/221/1/7,
        10.1088/0004-637X/775/1/49}. 
 A specifically dedicated RM campaign focused on 
active galactic nuclei  with H$\beta$  asymmetry {(Monitoring AGNs with H$\beta$ Asymmetry, MAHA)} has been running since 2017, which uses the Wyoming Infrared Observatory (WIRO) 2.3m telescope \citep{2018ApJ...869..142D,10.3847/1538-4357/abc2d2,Bao2022}. However, the observational data  are inconclusive, and further monitoring is needed.

 Thanks are given to  GRAVITY (General Relativity Analysis via VLT InTerferometrY) on board the Very Large Telescope Interferometer (VLTI) \citep{Hag12,10.1051/0004-6361/201730838} for bringing in  a new era of interferometry for high-spatial-resolution astronomy. GRAVITY operates in the K band, between 2.0 and 2.4 $\mu$m, interferometrically combining near-infrared  (NIR) light collected by four telescopes at the VLTI \citep{10.1051/0004-6361/201730838}. {It successfully  observed \object{3C 273} and the data obtained allowed the inference of the radius of its broad line region   
 \citep[BLRs;][]{10.1038/s41586-018-0731-9,10.1038/s41550-019-0979-5}, a $\sim 20\%$ error in its SMBH mass estimate, and cosmic distances \citep{10.1038/s41550-019-0979-5}.}
 The second source is \object{IRAS 09149-6206},  
 for which \citet{10.1051/0004-6361/201730838} measured  the size of the BLR ($\sim 0.075$ pc) and the mass of the central black hole ($\sim 10^{8} M_{\odot}$), while \object{NGC 3783} is the third \citep{Taro2021}. {The} GRAVITY instrument partially resolved the continuum hot dust emission {of eight AGNs,} with hot dust continuum sizes {ranging from} $0.3$ to $0.8$ mas \citep{10.1051/0004-6361/201936767}. {The} hot dust continuum of \object{NGC 1068} {was}  spatially resolved \citep{10.1051/0004-6361/201936255}, revealing a thin, ring-like structure with a radius of $\sim 0.24 $ pc.
 
 The proposed {GRAVITY/VLTI} upgrade, {known as} GRAVITY+, is intended {to broaden} interferometric frontiers toward $ K > 22\,$ mag \citep{mpe.mpg.de/7480772/GRAVITYplus_WhitePaper.pdf}, where detection of CB-SMBHs {is best accomplished} in collaboration with current, high-precision radial velocity (RV) \citep{10.3847/1538-4357/abc24f}  and quantitative spectroscopy programs \citep{10.3847/1538-4357/ab2e00, doi.org/10.3847/1538-4357/ab3c5e, 10.1038/s41550-019-0979-5,2020ApJS..247....3S}. GRAVITY+, by providing spatial information, will be the ultimate tool for securely establishing the binarity of candidates, which are predicted to be observed in the thousands in upcoming surveys.
 
 {
        Because of the uncertainty surrounding the  photometrically and spectroscopically selected candidates, various searches for more signatures  have been conducted and new detection methods are being developed.
        For example, the binary signature  may also be imprinted on
        the IR emission from the dust in the AGN \citep{10.1093/mnras/stx1269}.
       { Recently, \citet{10.3847/1538-4357/abc24f} developed a new technique to identify CB-SMBHs
with circular orbits ($e = 0$) based on astrometric signatures observed by GRAVITY+
that are a consequence of the morphology and evolution of hot dust emission in the
system.}

}
 {With the aid of GRAVITY+,  high-precision astrometry it will be possible to further probe {\eCBSMBH}  candidates selected  from Doppler-shifted emission-line  surveys. 
        This spectroscopic method detects
        binaries with longer periods of at least a few decades \citep{2022MNRAS.510.5929C}.
        It is commonly assumed that these two indirect  detection methods require observational time baselines exceeding the orbital period to produce positive results.  
 }

 {In this work, we  simulate synthetic and incomplete  astrometric and radial velocity observations of {\eCBSMBH}s   to investigate the effect of eccentricity on their astrometric and radial velocity signatures, the possibility of their detection,  and recovery of basic orbital elements.} Our technique differs from that of \citet{10.3847/1538-4357/abc24f} in that we used a greater parameter range (including {\eCBSMBH}\, eccentricity) and we considered a realistic and unfavourable percentage of the \eCBSMBH\, orbit covered by observations $(5\%-10\%)$. 
 { In a set of simulated astrometric and 
{radial velocity} (RV) observations} covering only $5\%-10\%$ of {a} whole orbital period of {the source}\,(which we refer to as the `interferometric gap'), we {illustrate the} Bayesian method as the plausible solution to this {issue}.
 Bayesian inference {is used} to combine the two sets of data, and Markov Chain Monte Carlo (MCMC) is applied  to produce random samples from a distribution of the orbital parameters based on the simulated observations
 \citep{10.1063/1.1699114, 10.1093/biomet/57.1.97, 10.7717/peerj-cs.55}.
 
 {
        The structure of the article is as follows. Section  \ref{sec:modelling}  presents our  {\eCBSMBH}  model, which includes astrometric and radial velocity data.
        In Section \ref{sec:detectability}, we first discuss the detectability of {\eCBSMBH s} in general, as assessed by robust astrometric signature amplitudes.
        Section \ref{sec:detect1} outlines detectability based on the photo-centre offset generated by the intersection of the secondary SMBH dust ring and 
{circumbinary disc} (CBD). Section  \ref{sec:detect2} highlights detectability in the limit of binary eccentricity, which influences orbital shape.
        Section \ref{sec:results} displays the results of the Bayesian procedure for orbital parameter recovery from synthetic  multi-data  sets (joint astrometric and radial velocity).
        Section \ref{sec:discussion} discusses {\eCBSMBH} detectability refinements based on the variation of $q, f_{\rm orb}$ parameters,  the possibility
        of obtaining orbital eccentricity from radial velocity and acceleration data, and refinement of {\eCBSMBH} detectability in contrast to CBD emission phenomena. We finish this section by introducing the Joint Spectroastrometry and Reverberation Mapping (SARM) approach, which can be used for binary detection refinement via follow-up or as an independent binary-detection tool.
        In Section \ref{sec:issue}, we describe  the limitations of model assumptions and the challenges in radial velocity and centroid measurements. Section \ref{sec:prospect} shows  a possible approach  for determining the angular position of the photo-centre at the emission line  for future   successors  of the GRAVITY+ instrument. In addition, the overall expectation of {\eCBSMBH}  
{gravitational wave}(GW) measurements is outlined. In Section 8 we present our conclusions with some closing remarks.
 }

\section{{\eCBSMBH} model settings } \label{sec:modelling}
\subsection{{Overview} of accretion on to  CB-SMBHs  }

Here we {briefly explain the technique we use} for multi-data survey {modelling} of the \eCBSMBH, {which includes} astrometric measurements and RV observations, {as well as} the anticipated CBD and {hot}-dust ring characteristics of \eCBSMBH.
{The framework of our model is based on general CB-SMBH features deduced from theoretical studies. According to hydrodynamic simulations, an SMBHB opens a cavity in the surrounding gas, forming a circumbinary accretion disc \citep{ 10.1086/173679, 10.1086/523869, 10.1088/0004-637X/783/2/134}. As gaseous streams enter the cavity, some of the  matter becomes attached to the SMBHs, and at least one and possibly both of the SMBHs will acquire its own accretion disc and appear as an AGN \citep[e.g.][]{10.1088/0004-637X/783/2/134}. Furthermore, the accretion rate  is higher on the component with the lowest mass  in unequal-mass CB-SMBHs \citep{10.1086/173679, 10.1093/pasj/59.2.427, 10.1111/j.1365-2966.2011.18927.x, 10.1088/0004-637X/783/2/134}, potentially making it more luminous than the primary \citep{Bogd21,  10.3847/1538-4357/abe386}. Additionally, mass accretion is higher
        onto the secondary SMBH because it moves closer to the edge of
        the CBD.
        While the inner minidiscs are assumed to be responsible for the majority of the UV and X-ray emission \citep{10.3847/1538-4357/aad8b4, 10.1111/j.1365-2966.2011.20097.x}, the circumbinary disc is expected to be responsible for the optical and IR emission \citep{10.1038/nature15262}. 
        Simulations demonstrate that a dense and relatively cold circumbinary disc can transfer angular momentum whilst also being radiatively efficient and similar to discs that power AGNs, generating a luminous {electromagnetic} (EM) signal independently of GW emission during inspiraling \citep[see][and references therein]{10.1093/mnras/stab1022}. Many simulation results \citep[see e.g.][]{10.1093/mnras/sty423, 10.3847/1538-4357/aaf867}  show that after gap  formation during the binary--disc interaction, unique observable signatures of the continuum emission could be observed
        \citep{10.1088/0004-637X/761/2/90, 10.1093/mnras/stab1022}.}

{
        The current premise for tracking binary SMBHs using GRAVITY is based on the relative astrometry between the BLR  of an accreting  black hole and hot dust in the surrounding circumbinary disc \citep{10.3847/1538-4357/abc24f}.
        We assume that the secondary SMBH has a higher accretion rate and is more luminous than the primary \citep[see e.g.][]{10.3847/1538-4357/abc24f,10.3847/1538-4357/abe386}, which is broadly referred to hereafter as the `active SMBH'
        .\\}
{
        As the main unknown in this setup is where the NIR continuous emission comes from and how it evolves over the binary motion,  we look at two scenarios as prescribed by \citep{10.3847/1538-4357/abc24f}. First, hot dust emission is stationary (e.g. uniform or asymmetric around the circumbinary disc), allowing the relative astrometry of the  BLR to be exploited in order to compute the orbit of the secondary. Second, hot dust is assumed to originate outside the binary and at the sublimation radius of the  secondary. The probable emission locations are then distributed along a circle of radius $\Rsub$ centred on the secondary position. When the sublimation radius intersects the circumbinary disc, it is anticipated that hot dust emission occurs with equal intensity all the way around the portion of the circle intersected by circumbinary disc. As assumed by \citep{10.3847/1538-4357/abc24f}, when the sublimation radius is less than the distance from the secondary to the edge of the circumbinary disc, it is anticipated that a tiny region (e.g. in an accretion stream) at a distance of $\Rsub$ will develop and emit hot dust instead.
}

\subsection{Number of possible {\eCBSMBH} systems within reach of GRAVITY+ } 
{
        \indent To better gauge the frequency of  {\eCBSMBH s } 
        amongst the population of SMBHBs, we  estimate  their frequency distribution ($f_{\rm\eCBSMBH}$) by   integration of a differential fraction within an arbitrary range of eccentric binary masses $M$, eccentricities ($e$), and periods ($P$):
        \begin{equation}
        f_{\rm\eCBSMBH}=\mathcal{C}\iiint p_{\rm active}p(M)p(e)p(P) dM\, de\, dP
        \label{eq:freqecb}
        ,\end{equation}
        \noindent where  $\mathcal{C}$ is the normalisation constant dependent on the sample;   $p(M),p(P)$, and $p(e)$   are the  distributions of {\eCBSMBH} mass,  eccentricity, and period; and  $p_{\rm active}\sim 0.5$ is an approximate probability that the secondary is active and more luminous. For probabilities, we use approximated functional power law forms $p(M)=M^{-1.1}, p(P)=P^{-1}$, and $p(e)=e$, which  are simple   because of the unknown sample of binaries in the range of parameters of interest.
        An estimate of the number of {\eCBSMBH s} at a given distance ($z<0.3)$ whose astrometric signal could be detected by GRAVITY with  CI 95\%  is then given by $N_{D}\sim 0.95f_{\rm\eCBSMBH} N_{\rm GRAVITY}$, where $N_{\rm GRAVITY}$ is the total number of AGNs detected by GRAVITY within a sphere
        of radius $z<0.3$, while $f_{\rm\eCBSMBH}$\, is calculated by integrating over a specific range of masses, periods, and eccentricities $(M\in[10^{7},10^{10}]M{\odot},P\in[5,100]yr,e\in[0.4,0.6]) $. If $N_{\rm GRAVITY}$ is of the order of a few hundred ($<500$) within $z<0.3$ \citep{mpe.mpg.de/7480772/GRAVITYplus_WhitePaper.pdf} then the number $N_{D} < 69 \mathcal{C}$. We see that if $\mathcal{C}\sim 0.05-0.2$, the number of \eCBSMBH s\, is $4-13$ which is comparable to the established set of 10 circular GRAVITY+ targets.
        In addition to the above blind estimate, we can  calculate the number of SMBHBs that can be detected by GRAVITY+ up to $z\sim 3$ using {the estimated number} of SMBHBs  per $\log z$ \citep{10.1103/PhysRevD.100.103016} and assuming   the fraction of
        CB-SMBHs in local bright AGNs is $f_{b}\sim 10^{-3}$\citep{10.1088/0004-637X/703/1/L86}}:
{
        \begin{equation}
        \frac{dN}{d\log z}=f_{b}\,4\pi 
        \left(\frac{d^{2} V}{dz d\Omega}\right)\Psi(L)
        \min\left(\frac{t_{\rm res}}{t_{l}},1\right) (1+e^{-2W}),
        \label{eq:freqecb1}
        \end{equation}
        where
 ${d^{2} V}/{dz d\Omega}$ is the co-moving volume per redshift and solid angle ($\Omega\sim 4\pi$),
        
        \begin{equation}
        \Psi(L)=\frac{\phi_{0}}{({L}/{L_{0}})^{\gamma_{1}}+({L}/{L_{0}})^{\gamma_{2}  } }
        \end{equation}
        is the quasar luminosity function \citep[see][parameters are given in the last row of their Table 3]{10.1086/509629}, 
        \begin{equation}
        t_{\rm res}=\frac{20}{256}\Big(\frac{P}{2\pi}\Big)^{8/3}\Big(\frac{GM}{c^3}\Big)^{-5/3} q^{-1}_{s}*(1-e^{2})^{3.5}   
        \end{equation}
        is the residence time of the binary due to gravitational wave emission, $t_l\sim 10^7$ yr is the approximate AGN lifetime, $W=10yr-P_{\rm min}$ where 10 yr is the approximate GRAVITY mission, and we adopt a flat probability of eccentricity distribution $p(e)\sim e$. For simplicity, we assume that, at larger redshifts, we expect brighter and more massive sources and $e=0.5$. With these optimistic settings, we calculate that seven eccentric binaries could be detected in the sphere $z<3$.
        {As can be seen  from Equations \ref{eq:freqecb}-\ref{eq:freqecb1}, the frequency distribution of {\eCBSMBH s} is reliant on binary system characteristics;  in the following section, we therefore describe the models that are used in this study.}
}

\begin{figure*}
\centering
        \includegraphics[trim=15cm 20cm 20cm 30cm, clip=true, width=0.45\textwidth]{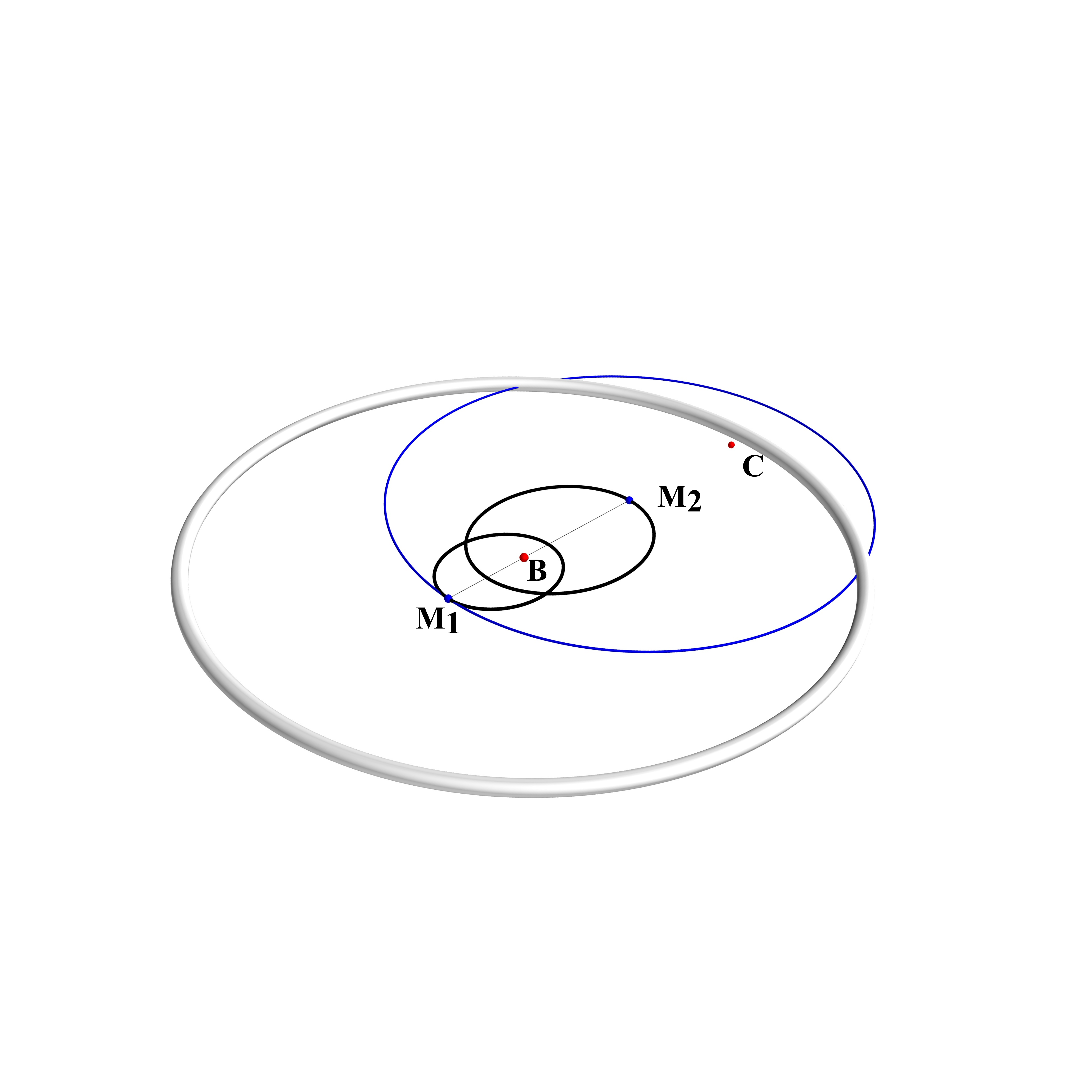}
        \includegraphics[width=0.45\textwidth]{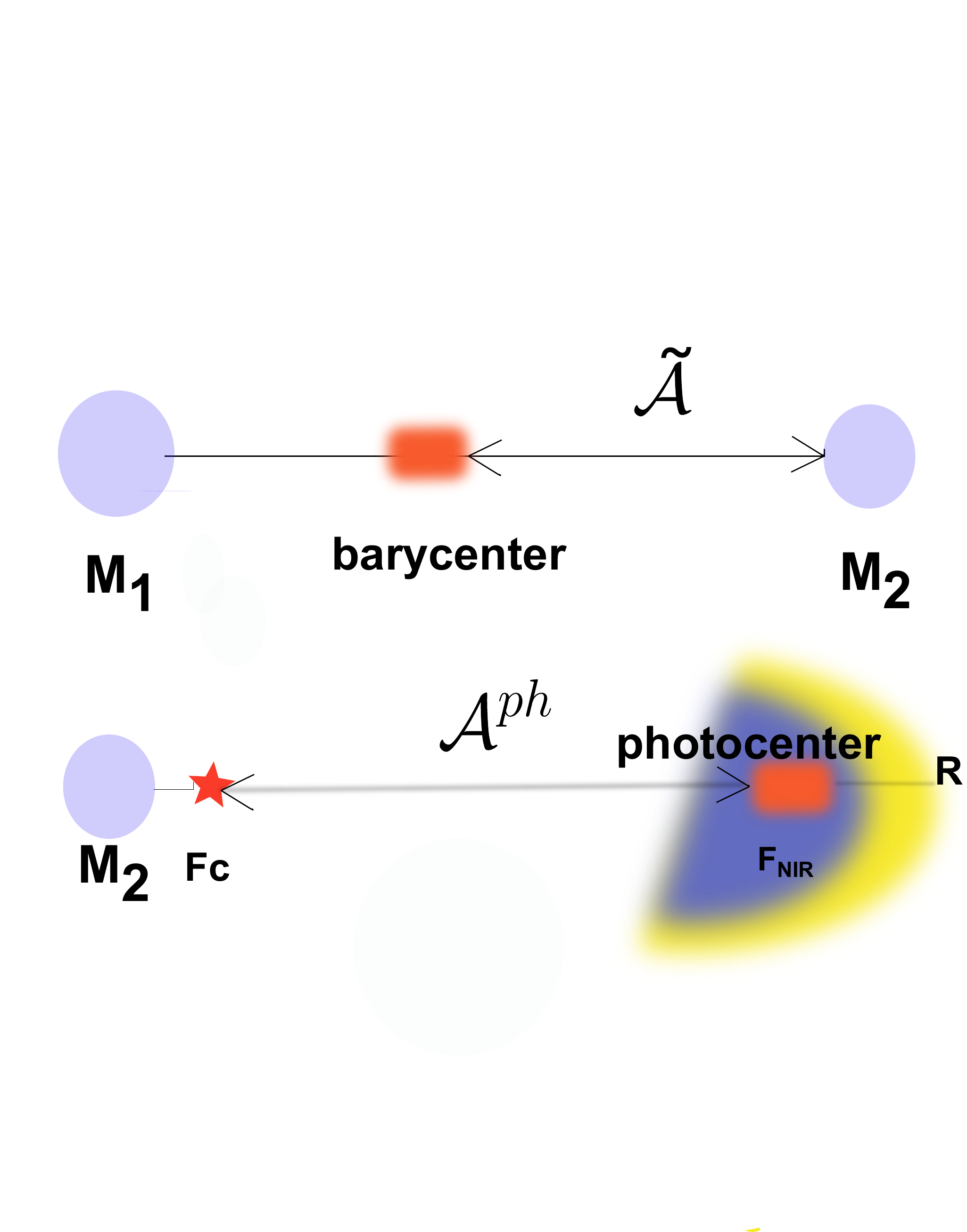}
        \caption{Evolving model for the continuum hot dust emission of {\eCBSMBH}. \textit{Left}: {Computer simulation snapshot illustrating the \eCBSMBH\, configuration at the time instance of pericentre passage.} $M_{1}$ and $M_2$ are SMBH loci, black ellipses are their  orbits, $B$ is the barycentre of the system, the blue circle is the sublimation ring bound to the $M_2$, the grey ring is the CBD, $C$ is the centroid of the arc of the dust ring outside the CBD and the photo-centre of the continuum in GRAVITY's $K$ band.
                The masses of {the SMBHs are}
                $M_{1}= 6\cdot 10^{7} M{\odot}$ and  $ M_{2}=4\cdot 10^{7} M{\odot}$, \eCBSMBH\, eccentricity is $e=0.5$, and other orbital parameters {are} $\Omega_{1}=0.1^{\circ}, \omega_{1}=0.1^{\circ}, \Omega_{2}=180.1^{\circ}$, and $\omega_{2}=180.1^{\circ}$. { See
the main text for a  description of the coordinate system}.
                \textit{Right} Upper: Schematic of the  barycentric  photo-centre displacement $\tilde{\mathcal{A}}$ used in astrometry to detect a  mass $M_2$ gravitationally bound in a two-body system with mass $M_1$. Bottom: Nearby $M_2$ originates optical continuum $F_c$, which is {then} reprocessed by a dusty torus (yellow circle) into NIR emission  $F_{NIR}$ in {the} K band (2.2$\mu$m), {commonly known} as `dust continuum'. {Because} the dust ring (yellow circle segment) and CBD (blue circle segment) {intersect}, the photo-centre will be the centroid of the arc of the dust ring (yellow circle). The distance between the highest point of the dust-ring arc and $M_2$ is  {denoted by the symbol}  $h$. 
                $\mathcal{A}^{ph}$ is the angular displacement
 of the photo-centre with respect to $M_2$. }
        \label{model1}
\end{figure*}

\subsection{Setup of an astrometric model and an RV data model}\label{sec:astr}

We {consider} the \eCBSMBH\, model {to be a} two-body system of SMBHs,  { such that $M_{1}>M_{2}$} (see the left panel in Figure \ref{model1}). 
The formalism is {discussed} briefly {below}, and further {information} may be found in  \citet{10.1051/0004-6361/201936398}. 
{The true motion of the two components around the barycentre of the system (B)
        lies in the relative orbital plane of the binary. This
        is called a coplanar SMBHB system. The common orbital plane is
        set as the reference plane of the barycentric frame\footnote{The orbital plane of a more massive SMBH might be employed for the perturbed non-coplanar system.}.
        The common binary
        orbital plane is perpendicular to the vector of the binary orbital angular momentum, which is fixed to  the barycentre. This vector serves as the  $Z$-axis of the barycentric frame, whilst the barycentre B serves as the origin of the frame.
        The reference plane is spanned by the $X$-axis  (aligned with the 
        { semimajor axis of binary relative orbit}} 
        and pointing from the barycentre to the pericentre)  and the Y-axis (perpendicular to both the 
        $X$- and $Z$-axis, making a righthanded triad).
        The primary and secondary orbits could be orientated in any direction to the observer.
\begin{table}
        \caption{Seven parameters  required to describe a Keplerian orbit of {\eCBSMBH} in three dimensions.}
        \centering
        \setlength{\tabcolsep}{1pt}
        \begin{tabular}{lccc}
                \hline\hline
{Parameter} & {Units}&{Name}& {fiducial range}\\  
                \hline
$a$ & ld $\vee$ pc&semimajor axis & $[0,\infty)$ \\
$e$ & -&eccentricity &$[0,1]$ \\
$P$ & yr $\vee$ days&orbital period &$(0,+\infty)$ \\
$\omega $& $^{\circ}$&argument of periastron&$[0, 360]$\\
$i $& $^{\circ}$&inclination&$[-90, 90]$\\
$\Omega $& $^{\circ}$&angle of ascending node&$[0, 360]$\\
$T_{0}$ & days $\vee$ yr&time of periastron passage & $[0,\infty)$\\
                        \hline
        \end{tabular}
        \label{deforbit}
\end{table}

{Naturally, dynamical parameters fully describe the SMBH position relative to the barycentre (see Table \ref{deforbit}).
        The apparent relative orbit is that of the secondary around the primary projected on the sky plane, and it can be determined from measurements of the relative position of the components obtained through astrometric imaging or interferometric observations.
        The projected spatial motion of the binary components is described using the reference frame centered on the primary component or barycentre and two axes in the plane tangent to the celestial sphere 
        \citep{10.1051/0004-6361/201220454}:  the $x$-axis points north, while the $y$-axis points east. The $z$-axis runs parallel to the line of sight and points in the direction of rising radial velocities (positive radial velocities).}

The transformations could be {represented} in the vectors $\mathbf{P}$ and $\mathbf{Q}$ (or equivalently Thiele-Innes parameters) instead of {using} cosine and sine terms of rotations. 
It is also {feasible} to include any observer position \citep{10.1051/0004-6361/201936398}.  
The vector of relative position $\mathbf{r}(t)=[x(t),y(t),z(t)]$ of an SMBH 
with {regard} to the barycentre of the  {system} can be {expressed} in compact form as
\begin{equation}
\mathbf{r}(t)=\mathbf{r(0)}+\partial_{t}\mathbf{r}(0)t +\mathbf{P}[\cos E(t)-e]+\mathbf{Q}\sqrt{1-e^2}\sin E(t),
\label{eq:compose}
\end{equation}
where $E(t)$ is {the }eccentric anomaly {determined} from the Kepler equation $E(t)-e\sin E(t)=2\pi n (t-t_{0}), n=P^{-1}$, {and} $t_{0}$ {is set to} zero  for simplicity. 
The inertial frame {is defined by} constant vectors of position
$\mathbf{r}(0)$ and velocity $\partial_{t}\mathbf{r}(0)$, { which are }set to zero for simplicity. $\mathbf{P}$ and $\mathbf{Q}$, the auxiliary vectors, {are defined as follows}:
\begin{eqnarray}\label{thiel1}
\mathbf{P} &=& {a_{\bullet}} \left[\mathbf{p}\cos(\omega)+\mathbf{q} \sin(\omega)\right],\\
\label{thiel2}
\mathbf{Q} &=& {a_{\bullet}}\left[-\mathbf{p}\sin(\omega)\ + \mathbf{q}\cos(\omega)\right],\\
\label{thiel3}
\mathbf{p}&=&(\sin \Omega,\cos \Omega,0),\\
\label{thiel4}
\mathbf{q} & =&\left(\mathcal{I}\cos{\Omega},-\mathcal{I}\sin\Omega,\sin i\right).
\end{eqnarray}
The data matching {the third coordinate of body position} ( $z(t)$) cannot be {obtained}, but the radial velocity ($\dot{z}(t)$), which is  the time derivative of $z(t)$,
{may be measured as follows}:
\begin{eqnarray}
V_{\rm rad}=\dot{z}(t)&=&\frac{2\pi P} {[1-\cos E(t)]}\left[ C \sin E(t)-H\sqrt{1-e^{2}}\cos E(t)\right],\\
\label{radvel11}
C&=&a_{\bullet}\sin(i)\sin(\omega),\\
H&=&a_{\bullet}\sin(i)\cos(\omega).
\label{radvel}
\end{eqnarray}
Finally, the orbital position (Equation  \ref{eq:compose}) and the radial velocity (Equation \ref{radvel}) {may be expressed} in compact form.
{Recognizing} that the auxiliary vector components 
$\mathbf{P}=
\begin{pmatrix}B\\A\\C\end{pmatrix}$ and $\mathbf{Q}=\begin{pmatrix}G\\F\\H\end{pmatrix}$ are the Thiele–Innes (TI) elements,  { we may use}   Equations  \ref{eq:compose} and  \ref{radvel} { to calculate SMBH positions  on the sky plane from:  } 
\begin{eqnarray}
x=x_{0}+BX+GY,
\label{compose11}
\\
y=y_{0}+AX+FY,
\label{compose12}
\\
\dot{z}=C\dot{X}+H\dot{Y},
\label{compose13}
\end{eqnarray}
where 
$$X=\cos E -e, Y=\sqrt{1-e^{2}}\sin E,$$
$$\dot{X}=\frac{2\pi P} {[1-\cos E(t)]} \sin E(t), 
\dot{Y}=\frac{2\pi P} {[1-\cos E(t)]}\sqrt{1-e^{2}}\cos E(t).$$
The barycentre  {coordinates} $(x_{0},y_{0},z_{0})$ can be included into fitting parameters \citep{10.1515/astro-2017-0019}; however, {in this case} we assume the relative position of the secondary with respect to the primary, and {therefore} they are set to zero.

{The preceding} sets of equations should be {modified}   for the semi-major axis of the apparent { ellipse}  ($\tilde{a}$), {which} should replace the barycentric semi-major axis of either component ($a_{\bullet}$). 
{The} Newtonian generalisation of Kepler's third law {yields} a semi-major axis of the apparent ellipse  ($\tilde{a}$): 
\begin{equation}
\tilde{a} = \left[\frac{P^{2}\, G\, (M_{1}+M_{2})}{4\pi^{2}}\right]^{1/3}.
\end{equation}

{As a result}, the orbital parameters in Equations \ref{compose11}-\ref{compose13} { match} the relative orbit of the  {secondary}. {It is worth noting} that
$X$ and $Y$ are the displacement in the true plane.
The measured separations and position angles ($\rho,\phi$) of a secondary at time $t$ are {linked} to the projected quantities $(x, y)$ by the superficial {equations} $x=D \rho \cos\phi, y=D \rho\sin\phi,$ where $D$ is the distance to the object, and $\phi$ is called the position angle  
(PA).

{Because} the astrometric data, $\mathbf{\Xi}(t)=[\xi_{x(t)}=\rho\cos\phi,\xi_{y(t)}=\rho\sin\phi],$ are orbital motions projected in the tangential plane and radial velocity  $\dot{z}(t)$  data are radial projections,  we {may}  combine these sets  into a multi-data ensemble:

\begin{equation}
\mathcal{M}(t)=(\mathbf{\Xi}(t),\dot{z})=
([\xi_{x}(t),\xi_{y}(t)],\dot{z})= (D^{-1}\times[x(t),y(t)],\dot{z}(t)).
\label{eq:compmodel}
\end{equation}
\noindent
A complete description of the {binary} system
{contains, in addition to} orbital elements,  the masses $M_{1}$, $M_{2}$, and distance. We assume that both masses and distances are known.

To be fitted to the recorded data for epoch $t$,{ both the models for projected line-of-sight velocity and radial velocity and the projected locations in the plane of the sky}  ($\mathbf{\Xi}(t)$), {known as astrometry \citep{10.3847/1538-3881/aa5e4a}}, require anomalies (mean $M$, eccentric $E,$ and true $f$). Thus, the anomalies are calculated first \citep{10.1051/0004-6361/201936398},  followed by radial velocity, and {then} the Thiele–Innes equations (Equations \ref{thiel1}- \ref{compose13}) {are used to estimate} the relative positions.

\subsection{Relevance of SMBHB eccentricity}

{ We then addressed the  extensive theoretical evidence of the relevance of SMBHB eccentricity as a general picture of SMBH binarity.}
Studies of the development of the orbital eccentricity of binary SMBHs {contained in} circumbinary discs {suggest} that the exchange of angular momentum within the system {causes} a continuous increase in binary eccentricity in the range 0.6-0.8. \citep{10.1086/497108,10.1111/j.1365-2966.2008.14147.x,10.1111/j.1365-2966.2011.18927.x}. 
{However, we focus on \eCBSMBH s\,  with eccentricity $\sim 0.5$ \citep{ 
                10.3847/0004-637X/828/2/68, 10.3847/1538-4357/aaeff0}, for which the inner edge radius of the circumbinary disc is  $\sim 2.5a$ \citep{10.1093/pasj/65.4.86}. Because an eccentricity of $0.5$ is less than the Laplace limit, the typical power series in the solution to the traditional Kepler equation converges \citep{Moulton70, 10.1103/PhysRevD.102.084042}.}

Furthermore, even in the late inspiral phase, SMBHBs formed in gas-rich galaxy mergers may retain substantial eccentricities \citep{10.1086/497108,10.1111/j.1365-2966.2008.14147.x}.
Additionally, N-body simulations of large galaxy mergers produce SMBHBs on eccentric orbits as a result of star interactions \citep[see e.g.][]{ 10.1088/0004-637X/695/1/455,10.1088/0004-637X/749/2/147,10.1088/0004-637X/773/2/100}. Also, the Kozai–Lidov oscillation \citep{10.1086/378794} might lead to eccentric mergers, in which a distant third object perturbs the binary orbital motion.

N-body simulations of  significantly non-spherical major mergers \citep{10.1088/0004-637X/732/2/89, 10.1088/0004-637X/749/2/147} reveal that the coalescence times of SMBHBs are shorter than those expected in spherical models, whereas binary eccentricities stay high throughout the simulations. In these simulations, SMBHBs, for example, could evolve in merger remnants to very high eccentricities of  $\sim 0.8-0.99$ with coalescence times ranging from 1 to 1.5 Gyr. For steeper density profiles  of merging galaxies,  binary eccentricities are  in the $0.5-0.8$ range, although  the coalescence time is shorter ($0.6-0.8$ Gyr).  In very steep-profile galaxy mergers, SMBHBs with eccentricities of  $0.4-0.6$ and very short coalescence times of $\sim0.4 \mathrm{Gyr}$ are found  \citep{10.1088/0004-637X/732/2/89, 10.1088/0004-637X/749/2/147}.

Furthermore, numerical simulations indicate that the evolution of the orbital eccentricity of an SMBHB embedded in a circumbinary disc is independent of the mass ratio of the  system, but is reliant on the barycentric location ({$\mathcal{L}$} )\footnote{$\mathcal{L}=R_{t}/a$, where $R_t$ is the distance of the strongest torque on the binary as measured from the centre of mass, and $a$ is the semimajor axis of the binary.} of the inner edge of the disc \citep{10.3847/0004-637X/817/1/70}. For $2 <\mathcal{L}< 2.5$, binaries will converge to a critical eccentricity value of $0.55 < e^{c} < 0.79$. Binaries with initial eccentricities $e > e^{c}$ will pass through a steady decrease in eccentricity, whereas binaries with $e < e^{c}$ will show the increase \citep{10.3847/0004-637X/817/1/70}. {Also, numerical simulations of the interaction between an eccentric SMBHB and its circumbinary gas disc suggest that eccentricity can be at least 0.01 just a week before coalescence \citep{10.1016/S1384-1076(96)00012-7, 10.1016/S1384-1076(97)00039-0, 10.1086/497108}.}

\subsection{Physical features of circumbinary discs and hot-dust rings}\label{sec:physfeature}

The quasi-simultaneous NIR and optical spectroscopy study of the continuum around 1 $\mu$m in 23 well-known broad emission line AGNs \citep{10.1111/j.1365-2966.2011.18383.x} {reveals}  that the continuum around this wavelength {is dominated by}  two emission components, a hot-dust ring  and an accretion disc. The estimated average hot dust radii for most objects were {less than} 1 lyr, {with} more than half {falling} between a few tens of light days  and 200\,ld. 
{The alleged sublimation radius changes for some objects \citep{10.1088/0004-637X/700/2/L109} have now been questioned, and if anything, the minor variations are debatable \citep{10.1051/0004-6361/201117750, 10.1088/2041-8205/775/2/L36}.}

Our goal is to show the astrometric approach to studying {\eCBSMBH s} 
using the observing capabilities of ground-based facilities.
The best AGN targets with hot-dust emission for such surveys are those in the redshift range $0.1 < z < 1.2$. 
GRAVITY+ upgrades will increase the number of observable type 1 AGN to hundreds at $z < 0.3$, more than a hundred at $z >0.8-1$, and a  dozen 
{quasars} at $z > 2$ \citep{mpe.mpg.de/7480772/GRAVITYplus_WhitePaper.pdf}.
{ Another} assumption (also used in spectroscopic searches for CB-SMBHs by \citet{10.1088/0067-0049/201/2/23, 10.1093/mnras/stx452}) is that the flux in the broad emission line is dominated by the gas flow bound to the secondary SMBH. Several theoretical studies of SMBHBs surrounded by CBD have directly motivated {this notion}  \citep{10.1093/pasj/59.2.427, 10.1111/j.1365-2966.2008.14147.x}.

Based on {the above}, we facilitate our goal by assuming {the simplest model} in which{ hot-dust    continuum emission is stationary, tracking  the inner edge of the circumbinary disc, or  hot dust is assumed to form outside the binary and at the sublimation radius ($\Rsub$) of the secondary  \citep{10.3847/1538-4357/abc24f}}.
{  Furthermore,} the dust is optically thin to its IR emission.  The dust ring {is expected to} obscure the BLR for lines of sight close to the plane of the accretion disc  \citep{10.1111/j.1365-2966.2011.18383.x}.

{In addition to}  the large body of literature addressing the{} theoretical
aspects of CBDs,  growing experimental evidence {supports} the CBD concept \citep[see][and references therein]{10.1093/mnras/staa1985}. {In simulations,} \citet{10.1086/523869} {detected} small {values} of eccentricity and ellipticity\footnote{ellipticity is defined for a spheroid analogously to eccentricity for an ellipse.} of CBD,  both between $0.05$ and $0.15$ at CBD radii of around $2a$. 
The maximum of these two values  is reached at much smaller radii $\sim a$. {In the case of a} misaligned disc, the inner part of the CBD tends to align with the binary orbital plane,   {while} the outer part tends to retain {its} original state \citep[see][and references therein]{10.1088/1475-7516/2015/07/005}.{ As a result}, we assume {that the }CBD {is} circular and that {its} orbit is coplanar with the  orbits of the  SMBHs.

Hydrodynamic simulations of prograde binaries (corotating with CBD) {demonstrate} that accretion occurs {mostly} on the secondary,  which orbits closer to the inner edge of the CBD in unequal-mass binaries 
\citep{10.1086/173679, 10.1111/j.1365-2966.2011.18927.x, 10.1088/0004-637X/783/2/134} and eccentric binaries \citep{10.1111/j.1365-2966.2008.14147.x, 10.1093/pasj/59.2.427, 10.1093/pasj/65.4.86,10.1088/0004-637X/783/2/134}.
Based on simulations of galaxy mergers, we {analyse} binaries with masses $10^{7}-10^{10} M_{\odot}$  and mass ratios {of} $0.1\leq q\leq 1$ for which SMBHBs are more likely to form  
\citep{10.1088/0004-637X/729/2/85,10.1093/mnras/stw2452}. Then,  for the secondary SMBH,  the range of considered masses  is $10^{6}\leq M_{2}\leq 10^{9} M_{\odot}$.
{With} a binary mass of $10^{7}-10^{10}M_{\odot}$ and {an} orbital separation {of} $\sim 0.01$ pc, orbital periods {range from} several decades  to a few centuries.
{$\Rsub$ is associated to the secondary luminosity, which together with the black
        hole mass is linked to Eddington ratio, as follows:
        \begin{equation}
        \Rsub\sim 0.4\frac{\epsilon L_{\rm Edd} M_{2}}{10^{46}{\rm erg}{\rm s}^{-1}}
        \end{equation}
        in units of parsecs,  where $L_{\rm Edd}$ is the Eddington luminosity, and $\epsilon = 0.1$ is the assumed Eddington ratio of the secondary.   This relation is derived \citep[see][]{10.3847/1538-4357/abc24f} using scaling relations with luminosity \citep{10.1088/0004-637X/767/2/149} and NIR continuum \citep{10.1051/0004-6361/201936767}. }

Here we assume {a} circular  CBD centred at the barycentre (B) of the \eCBSMBH.
If the dust ring and CBD are coplanar, {they}  will intersect in two locations, meaning that{ the following holds true}:
\begin{equation}
R_{\rm CBD}-\Rsub<\|\overrightarrow{BM}_{\,2}\|<\Rsub+R_{\rm CBD},
\end{equation}
where $\Rsub$ and $R_{\rm CBD}$ are {the} radii of the sublimation surface {attached to the secondary} and CBD, {respectively},  {while} $\|\overrightarrow{BM}_{\,2}\|$ is not constant over time for an elliptical orbit.

Assuming  ranges $ 0.5a<\Rsub<2.5a $ \citep{10.3847/1538-4357/abc24f} and $
R_{\rm CBD}\sim 2.5a(1+e)\sim 2a(1+e) $ \citep[see][]{10.1093/mnras/stu194, 10.1051/0004-6361/202039368} \footnote{Also it is possible to set $R_{\rm CBD}\sim 2.75a$ \citep[see e.g.][]{10.1093/pasj/59.2.427}.}, the intersection condition is 
\begin{equation} 1.5a+2ae<\|\overrightarrow{BM}_{2}\|<4.5a+2ae.
\end{equation}
{We briefly digress {to explain the exceptional case} of circular CB-SMBHs, {for which} $\|\overrightarrow{BM}_{\,2}\|=\frac{a}{1+q}= \rm const$ holds,  in order to {highlight} that the intersection {requirements} can be written using  $\Rsub$:
        \begin{eqnarray}
        R_{\rm CBD} - \frac{a}{1+q} <\Rsub < R_{\rm CBD} + \frac{a}{1+q}\\
        \implies a\Big(\frac{1+2q}{1+q}\Big)< \Rsub <a\Big(\frac{3+2q}{1+q}\Big)
        .\end{eqnarray}}
{Clearly, if $q=1$, {then} the CBD and dust ring will intersect if $1.5a<\Rsub<2.5a$.}
However, if the planes of the CBD and dust ring are inclined \footnote{
        In analogy  with the misalignment between dust and gas rings in circumbinary planetary discs due  to differences in their precession profiles \citep{10.1093/mnras/stab2794}.
} and their densities are non-neglibile at the crossing, then  a  slab-like region would be created  with direction  $\mathbf{L}_{\rm sub}\times \mathbf{L}_{\rm CBD}$ {rather than a } point-like emission structure.

{ Furthermore,} if the dust ring and CBD are {both centred} in the barycentre of the \eCBSMBH, they will not intersect. Secondly, the dust ring {is always considered to be} centred on the emission source; therefore, if the secondary is active and producing ionising radiation, the dust ring will be centred on the secondary.
The intersection of the dust ring and CBD will {result in} an {irregularity region} defined by their arcs. 
{Because of the generated irregularity}, the photo-centre {location} will shift outside the CBD arc to the centroid of the dust-ring arc. 
The new position of {photo-centre} {will be referred to} as  the astrometric perturbation.{ Moreover},  if the dust-ring is {positioned} at a radial distance $R_{\rm dust}$ {from} the hot accretion disc, {the} dust-ring  will reprocess the UV/optical  to thermal NIR radiation with a characteristic time-delay of 
$\tau_{\rm dust} = R_{\rm dust}/c$. {For around two-dozen} Seyfert galaxies, reverberation lags between NIR 
(K{-band}, 2.2 $\mu m$) and optical (V{-band}, 0.55 $\mu$m) light curves are {reported}
\citep{10.1086/381364,  10.1086/499326,  10.1088/0004-637X/788/2/159, 10.1051/0004-6361/201525910}.

Dust  {in the vicinity of}  AGNs absorbs the UV/optical radiation from the accretion disc and re-emits in the IR. The dust sublimates at $\sim 1500$ K ,{resulting in} the hottest dust emission,{ which peaks} at $\sim 2 \mu$m. Even though {the} Wein tail {diminishes} exponentially, {part} of  the hot-dust emission will reach  optical wavebands, as {demonstrated} by  \citet{10.1088/0004-637X/711/1/461}. {According to} \citet{10.1088/2041-8205/784/1/L4}, the fractional contribution of the dust  in filters i, z, {and} y is { particularly} sensitive to the redshift {of the object}.  
{The} dust contribution to the y-band is $\sim10\%$ {up to redshift } $z \sim 0.1$, {but declines} to $\sim 5\%$ at redshift $z = 0.2$. {Consequently}, in the following sections, we {incorporate} the NIR dust emission {into the model}.

We assume that NIR emission  $F_{NIR}$ is a scaled version of the optical continuum $F_C$  (i.e. $F_{NIR}\sim F_{c}^{\beta}$) because the response of the IR emission to the driving time variability of the AGN UV/optical continuum {may} be {described} as the convolution of the UV/optical continuum with a transfer function \citep{10.3847/1538-4357/aa7687}. {Similar} relationships {can be seen} in {the} optical band \citep{10.1111/j.1365-2966.2005.09795.x}.
{ The} left plot of Figure \ref{model1} { shows a   3D visualisation (in Mayavi2) of results from running simulations of the full model with typical \eCBSMBH} values.

\section{\eCBSMBH\,  detectability}\label{sec:detectability}

{
        We derive analytic expressions for the detectability of {\eCBSMBH s} in astrometric data, while taking into account  some basic GRAVITY+ parameters.
        We first find a simple analytical relation for detectability 
        based on the photo-centre offset caused by the evolving hot dust emission model (subsection 3.1). We then quantify detectability based on
        the astrometric signal in the limit of binary  eccentricity as a main factor of orbital shape (Section 3.2). 
        Both approaches are related to the signal amplitude  which is
        an order-of-magnitude estimate of detectability.
}

\subsection{The detectability of    \eCBSMBH\, astrometric signal based on a hot-dust emission source}\label{sec:detect1}

{We can estimate whether the astrometric signature of  \eCBSMBH s\, is above the  detection threshold of the GRAVITY+ instrument, understanding} that a detailed insight {is dependent on the  physics of the target and equipment features
}. 
{Because} the secondary SMBH is active and bright enough to be observed, we {explore the} following definitions. The barycentric astrometric displacement of $M_2$, {ignoring} the  dust-ring and the {CBD,} { caused by a} companion with mass $M_1$ {is as follows} \citep[see e.g.][]{10.1051/0004-6361/201015861}:
\begin{equation}
\tilde{\mathcal{A}}=\frac{M_1}{M_{2}}\frac{a_1}{D}
,\end{equation}
\noindent where $a_1$ is the barycentre-to-$M_1$ distance, ${D}$ is  the observer-to-object distance (see upper right plot in Figure \ref{model1}), and ${a_1}/{D}$ is the angular separation of $M_1$.
Based on {our prior discussion of the physical properties of}  CBD and the hot-dust ring in section \ref{sec:physfeature}, the NIR emission {flux}  ($F_{\rm NIR}$) is a scaled version of the optical continuum $F_{\rm c}$ (at $5100$ \AA),  {as follows:} $F_{\rm NIR}\sim F_{\rm c}^{\beta}$.

{We employ astrometry here in order to achieve accuracy below the resolution of the  system}. The dust region {tied} to the secondary intersects with CBD (see the left plot in Figure \ref{model1}) over {a specific period} of time {and may serve the above purpose}.
{Given the fact} that the BLR detection limit of an AGN is on the order of $\sim 40\,\mu \mathrm{as}$ for GRAVITY and $10\, \mu \mathrm{as}$ for GRAVITY$+$,  
{NIR} interferometric observations could  be used to map out the binary orbit by measuring the photo-centre difference between a broad emission line and the hot-dust continuum, rather than by resolving hot-dust emission \citep{10.3847/1538-4357/abc24f}.

{Assuming} that the photo-centre displacement is {caused} by an  irregularity (arc of  dust-ring cut by the CBD) at a distance $h=|\mathbf{h}|\sim \Rsub$ (see the right bottom plot in Figure \ref{model1}), the {position of} the centroid of brightness is

\begin{equation}
\mathbf{C}=\frac{F_{\rm NIR}(\mathbf{r_2}+\mathbf{h})}{F_{\rm NIR}+F_{\rm c}}+\frac{F_{\rm c}\mathbf{r_2}}{F_{\rm NIR}+F_{\rm c}}.
\label{eq:nirphoto}
\end{equation}

{
        Despite the exponential decrease of the Wien tail, some contribution of hot-dust emission will reach optical wavebands \citep{10.1088/2041-8205/784/1/L4}. \citet{10.1088/0004-637X/711/1/461} detected a dust contribution in the I-band after estimating the colour variability of optical variability. As a result, optical emission is made up of contributions from two distinct emission regions.}
{According to} \citet {10.1086/509878}, the accretion disc component contributes   $15\%-30\%$ {of} the NIR flux in the H band and $15\%-25\%$ in the K band and {may} be {calculated}  using V-band emission {data} \citep[see][]{10.1086/509878, 10.1088/0004-637X/700/2/L109}.  
{Therefore, we} assume that  $F_{\rm c}$ was {determined} beforehand.

Simply  subtracting the term $\mathbf{r}_{2}$ from the left and right sides of Equation \ref{eq:nirphoto} {yields} the photo-centre displacement  with respect to the $M_2$,  { as seen below}:

\begin{equation}
\mathbf{C}-\mathbf{r_2}=\frac{F_{\rm{NIR}}\mathbf{h}}{F_{\rm{NIR}}+F_{\rm{c}}}.
\end{equation}

The photo-centre angular displacement will {then} be {determined using the following formula:}
\begin{equation}
|\mathcal{A}^{\rm{ph}}|=\left|\frac{\mathbf{C}-\mathbf{r_2}}{D}\right|\sim \left|\frac{F_{\rm{NIR}}}{F_{\rm{c}}}\frac{\mathbf{h}}{D}\right|,
\end{equation}

\noindent {with the assumption} that {the} $F_{\rm NIR}$ contribution is {substantially} smaller than {the} $F_{\rm c}$ {contribution}. {The quantity 
        $\mathcal{A}^{\rm{ph}}$ corresponds to $\Delta x$ in \citet{10.3847/1538-4357/abc24f}}.

The {total} photo-centre displacement $\Delta \alpha$ {is a superposition of} the barycentric dynamical astrometric displacement $\tilde{\mathcal{A}}$ and the photo-centric displacement (`perturbation')  $\mathcal{A}^{\rm ph}$ {produced by anomalies in the flux distribution of} the unresolved sublimation surface intersecting CBD.
{If the } scaling {relation} between the  optical continuum ($F_{\rm c}$) and NIR emission  ($F_{\rm NIR}$){ for the secondary SMBH}  is $F_{\rm NIR}\sim F_{\rm c}^{\beta}$ \citep{10.1111/j.1365-2966.2005.09795.x},  the  photo-centre  angular offset relative to the secondary is as follows:

\begin{equation}
\mathcal{A}^{\rm ph}=\frac{F_{\rm NIR}}{F_{\rm c}}\frac{h}{D}\sim F_{\rm NIR}^{1-{\alpha}}\frac{h}{D},
\label{eq:ratio}
\end{equation}
\noindent where $\alpha=1/\beta$.

{ For different photo-centre displacements, we first show the flux ratios of the NIR emission originating in the dust ring with respect to the optical continuum as a function of dust-ring diameter.}
{Using} the mean distance of {the ten} best GRAVITY+ {circular} targets ($\sim 700$ Mpc) {and} assuming late-Universe parameters $H_{0}=67.36\, \mathrm{km s^{-1}}, \Omega_{m}=0.3166,\Omega_{\Lambda}=0.6847$ \citep{10.1051/0004-6361/201833910}, we show in  Figure \ref{fig:gridgravity}(a)  that {as the} ratio $F_{\rm c}^{{\beta}-1}$ increases for a given $h$, {so does} $\mathcal{A}^{\rm ph}$. 
{Next, we  assess the detectability of such irregularities using the GRAVITY detection limit in K band to $\lambda_{lim}\sim1.95  \mu \mathrm{m}$ \citep{10.1051/0004-6361/201730838}.}

{To do so we} compute {a}  rough approximation  of 
\begin{equation}
F_{\rm NIR}/F_{\rm c}\propto \frac{1}{h^2} \frac{B(\lambda^{\rm ir}, T)}{B(\lambda_{\rm lim}, T)}
\end{equation}
{by} estimating flux in $F_{\rm NIR}$ {with a}  modified surface brightness description that {scales} with  $\sim 1/{h^2}$ in proportion to {the} Planck curve \citep{10.1051/0004-6361/201117367} and 
a continuum source whose brightness {is equal}  to the GRAVITY wavelength  detection threshold (i.e. setting astrometric observing wavelength $\lambda^{\rm ir}=2.2 \mu \mathrm{m}$  of irregularity region).\footnote{ \citet{10.1111/j.1365-2966.2011.18383.x}  show that the continuum around the rest frame  $1\mu$m comprises mainly two emission components, a hot-dust blackbody and an accretion disc, where 
the latter is dominant. For objects at $z\sim 0.95$, such a continuum would translate into the lower end of GRAVITY detection limits ($1.95-2.45)\mu$m.
} 

The blue curve in Figure \ref{fig:gridgravity}(a) shows the corresponding  lower limit for GRAVITY+ observing wavelength. 
GRAVITY+ {may detect} the astrometric signal of irregularity whose   $F_{\rm NIR}/F_{\rm c}$  is above  the blue curve.
Thermal emission and light scattering can be significant in the K-band \citep[see][]{10.1051/0004-6361:20040362}, and the lower limit detectability curve can vary.
Different mechanisms in the system {may broaden the parameter space} where the `irregular' region can be bright enough to {cause} a photo-centre shift {yet} remain unresolvable.

\subsection{The detectability of {\eCBSMBH}\, astrometric signal in the limit of eccentricity} \label{sec:detect2}

{The} astrometric signature of a given object decreases with increasing distance and is dependent on the signal-to-noise ratio (S/N). Here we present {an approximate} estimate of {the}  (S/N) for  {various} \eCBSMBH\, mass ratios and eccentricities.  {A method like this} will also {provide} an {estimate} of the distance from Earth {at} which an \eCBSMBH\, can be detected.

{To establish a generic relation for an \eCBSMBH}, {we consider} the astrometric signal of a circular binary, {which is} given by
\begin{equation}
{\mathcal{A}} = \left[\frac{P^{2}\, G\, (M_{1}+M_{2})}{4\pi^{2}}\right]^{1/3}\times D^{-1},
\end{equation}
\noindent where $D$ is the distance to the object.
However, when the radial velocity amplitudes of components ($K_{i}=\frac{2\pi Q \tilde{a}\sin i}{P\sqrt{1-e^2}}, Q=\frac{q}{1+q} \lor \frac{1}{1+q}, i=1,2$) are available,  the {astrometric} signature can be rewritten {as follows}:
\begin{equation}
\tilde{\mathcal{A}} = 
\left[\frac{P\, (K_{1}+K_{2}) \sqrt{1-e^{2}}}{2\pi\,\sin(i)} \right]\times D^{-1} \sim \left[\frac{P\, (K_{1}+K_{2}) \sqrt{1-e^{2}}}{2\pi} \right]\times D^{-1} .
\end{equation}

{We} can {simply} approximate the {relationship} between astrometric signals of circular and eccentric  binaries using the above terms, as follows:
\begin{equation}
\tilde{\mathcal{A}} \propto
\left\{
\begin{array}{ll}
\mathcal{A}& \mbox{if } e = 0 \\
\mathcal{A} \sqrt{1-e^2} & \mbox{if } e > 0.
\end{array}
\right.
\label{totalsignal}
\end{equation}

\noindent {However}, \cite{10.1051/0004-6361/201015861} {provide} more {stringent} constraints on the astrometric signal of an eccentric orbit.
{We anticipate} that  orbital period $P$, {binary} mass, and the S\textbackslash N
will be the primary parameters {influencing } \eCBSMBH\, detectability. {We} define the astrometric S/N ($\mathcal{S}$) based on standard data analysis, {which suggests that enhanced} S/N  {happens} with  increasing number of observations ($N$):

\begin{equation}
\mathcal{S}\propto\Gamma\,\frac{\tilde{\mathcal{A}}\times {N}^{\gamma}}{\sigma},
\label{eq:snrr}
\end{equation}
\noindent where $\sigma$ is the single epoch noise, and we {use the} factor $\Gamma\sim 1$ {to accommodate}  for the characteristic of the signal and power{ index}  $\gamma=0.5$ for simplicity.

{
        Eccentricity makes detection more challenging at short periods, because uneven sampling frequently results in poor phase coverage during  rapid pericentre passage. The width of pericentre passage is $\sim (1-e)^{2}P$ \citep{10.1111/j.1365-2966.2004.08275.x}, which means that for binaries with $P=1{\rm yr}, e=0.5$, observations should cover a small window of 3  months of periastron passage. 
        On the other hand, transition to a long-period regime occurs when $P\rightarrow \mathcal{T}/{(1-e)^{2}}$ \citep{10.1111/j.1365-2966.2004.08275.x}, which means that the  
        number of orbits observed is $N_{O}=\mathcal{T}/{P}\rightarrow(1-e)^{2}$. The final term should also represent the probability of a binary being in the correct phase (i.e. in the width of pericentre).  However, the enhanced velocity amplitude and acceleration near the periastron boost detectability in long-period objects. Specifically, this means that if viewed at the right phase, can have tracks that are incompatible with linear motion  even when the period is long. 
        Taking the above into account, we can update Equation \ref{eq:snrr} as follows:
        \begin{equation}
        \mathcal{S}\propto\Gamma\,\frac{\tilde{\mathcal{A}}\times {N}^{\gamma}N_{O} \,\iota}{\sigma},
        \label{eq:snrr1}
        \end{equation}
        \noindent where $N_O$ takes into account the arc of the observed binary orbit in the long-period regime and $\iota<1$ is related to {degradation} 
        of the observational cadence due to unpredicted situations. Thus, $N_{O}\cdot\iota$ gives the true coverage of the arc of orbit. When $\iota=1$, there is no unpredicted loss of observations.}

\begin{figure*}[htb]
    \begin{minipage}[]{.45\textwidth}
        \centering
        \includegraphics[width=\textwidth]{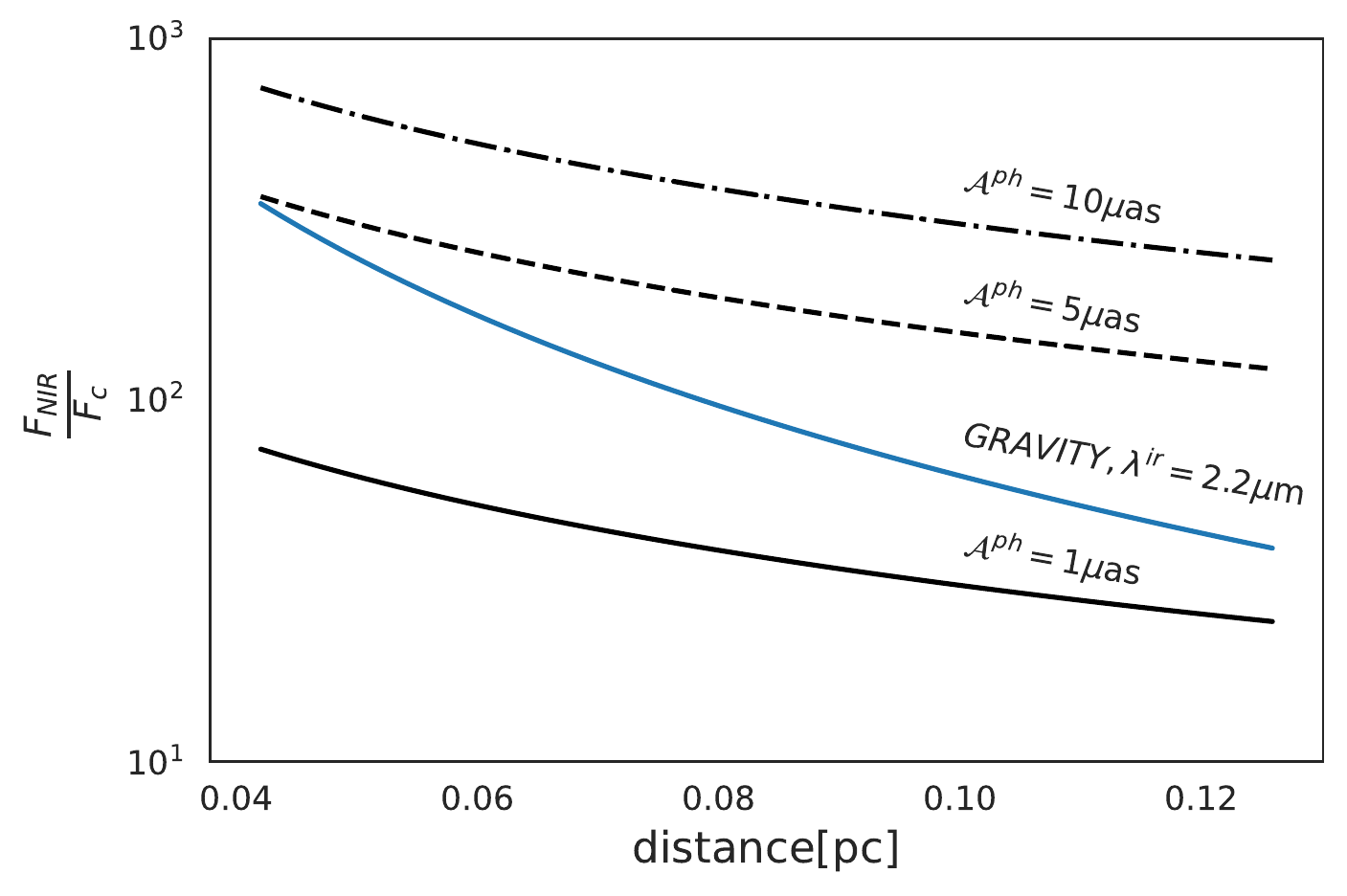}
        \subcaption{}
    \end{minipage}
    \hfill
    \begin{minipage}[]{.45\textwidth}
        \centering
        \includegraphics[width=\textwidth]{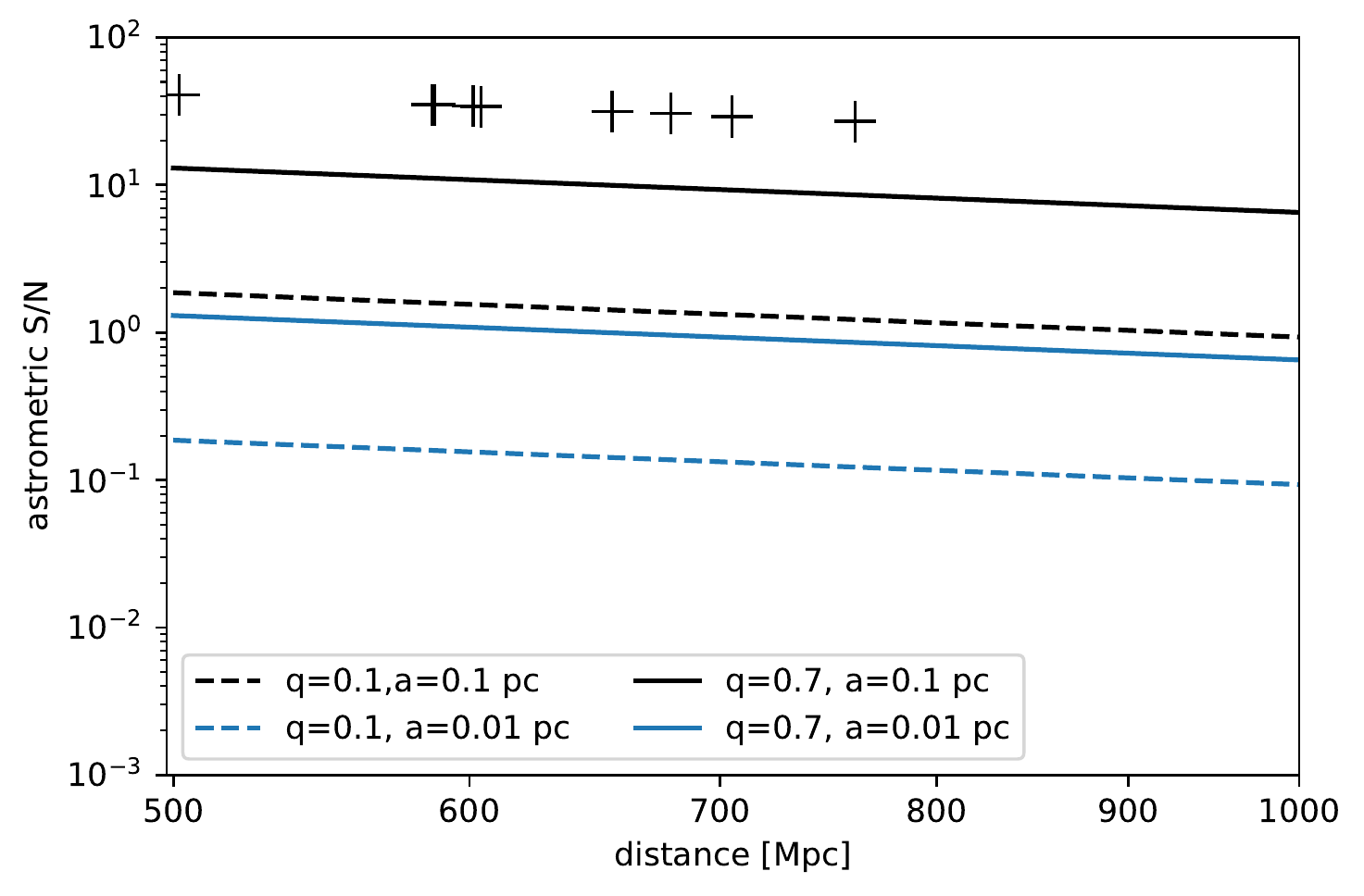}
        \subcaption{}
    \end{minipage}  
     \newline
    \hfill 1
      \begin{minipage}[]{.45\textwidth}
        \centering
        \includegraphics[width=\textwidth]{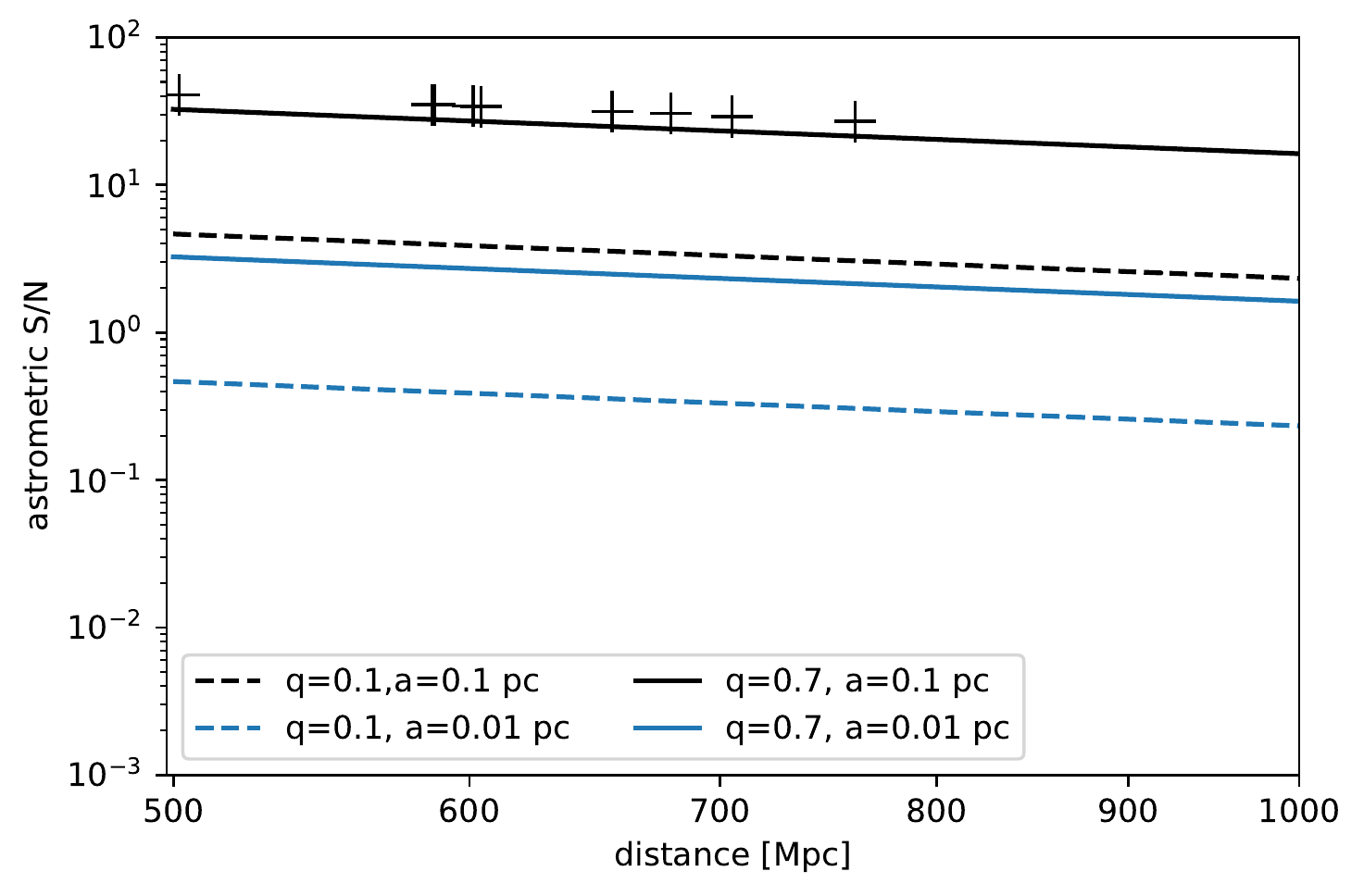}
        \subcaption{}
    \end{minipage} 
     \hfill
     \begin{minipage}[]{.45\textwidth}
        \centering
        \includegraphics[width=\textwidth]{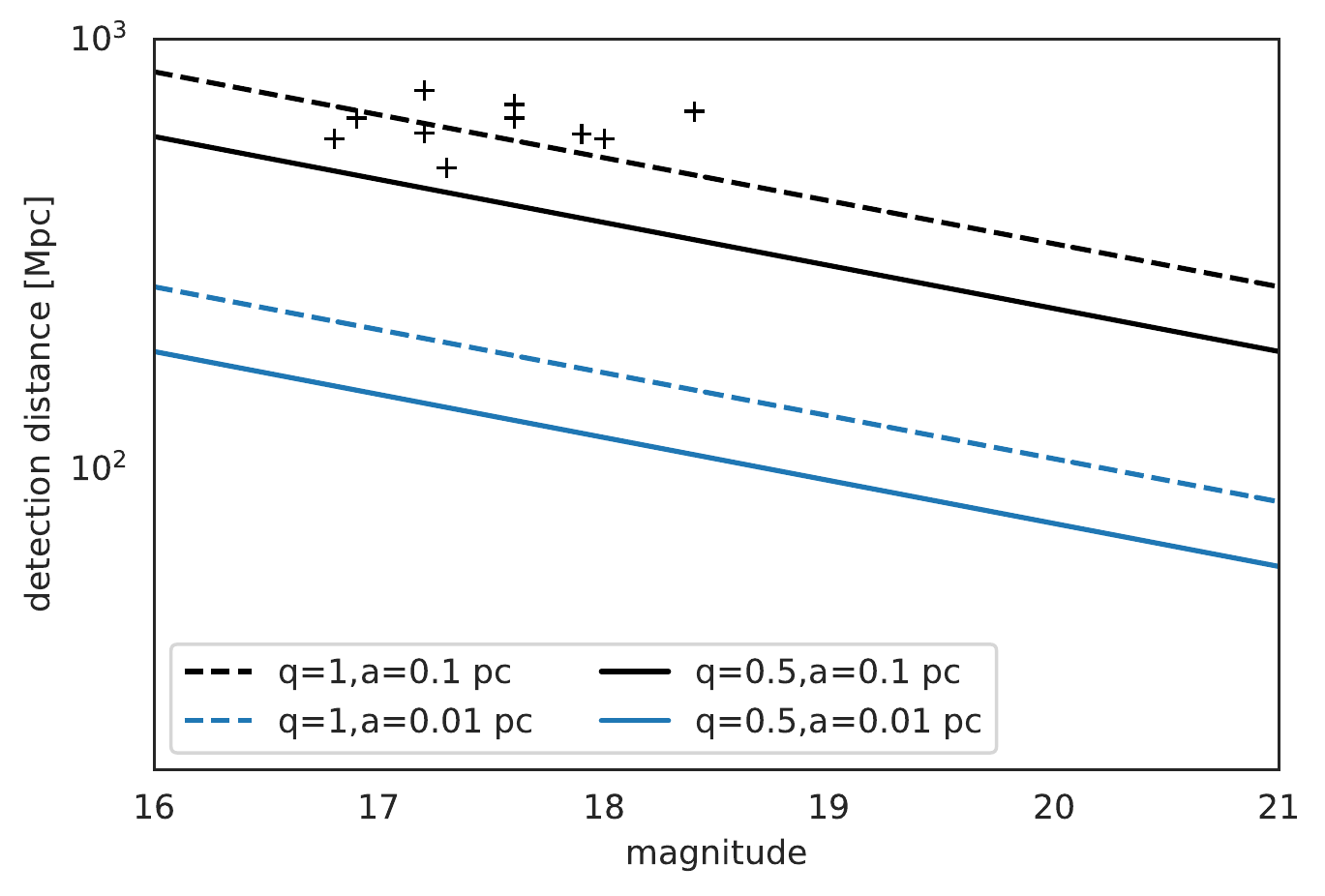}
        \subcaption{}
    \end{minipage}  
    \caption{{Various  aspects of \eCBSMBH\, {detectability} with GRAVITY +.} (a) 
    Flux ratios of the NIR emission originating in a dust ring with respect to the optical continuum {as a function of $h$ (dust-ring dimension)}, for different photo-centre displacements, shown as black lines. 
    The GRAVITY approximate detection limit {at  $\lambda_{lim}\sim 1.95  \mu \mathrm{m}$}  {in terms of} the flux ratio for dust emission arising at different distances and observed at $2.2 \mu \mathrm{as}$ is shown as a {blue} curve. {The shape of the blue curve depends on which power index of $h$ is chosen; here its value is set to  -2.} (b) Expected S/N for $N_{O}=0.25$ and $\iota=0.4$. (d) {Expected S/N for $N_{O}=0.25$ and $\iota=1$}. (c) Detection distance  for different \eCBSMBH\, parameters and eccentricity $e=0.5$. The crosses show  the ten best GRAVITY $+$ targets \citep[see][]{10.3847/1538-4357/abc24f}, assuming a circular
                orbit.}
                 \label{fig:gridgravity}
\end{figure*}

{We may evaluate some aspects of} \eCBSMBH\, {detectability} using Equation \ref{eq:snrr1}. This is illustrated in Figure  \ref{fig:gridgravity}(b) and (c), {which assume} \eCBSMBH s\, of {various}  mass ratios,  semi-major axes, $e=0.5, N_{O}=(1-0.5)\cdot{2}=0.25, \iota=\{0.4,1\},$ and $\sigma\sim 6 \mu \mathrm{as}$.
We optimistically expect  the  GRAVITY+ error in each coordinate {to be about} half of the present accuracy of GRAVITY in each coordinate $\sigma\sim (9.5/2) \mu$as \citep{10.1051/0004-6361/201423940}, such that the combined error measurement of both coordinates is $\sigma \sim \sqrt{2\cdot(4 \mu \mathrm{as})^{2}}$.

{The S/N} can be subjected to a given threshold, such as $1\lesssim\mathcal{S}$ {(i.e. the binary motion dominates
        over the error)}; hence, Figure  \ref{fig:gridgravity} {(b) and (c)  {provide} the approximate S/N needed to detect   \eCBSMBH s\, }. 
As {indicated} by the overplotted GRAVITY+ targets, 
the S/N of objects in circular orbits is expected to be higher.{ For observation loss of $\iota=0.4$, implying 10\% observational coverage of pericentre width, it is expected that {\eCBSMBH s} with $q=0.1$ at a mutual distance of 0.01 pc would be impossible to detect (see subplots (b) and (c)). However, S/N is approximately two or three times higher when there is no degradation in observational cadence ($\iota=1$, subplot (c)). These estimates support the detectability of {\eCBSMBH s}, which is discussed throughout the text.}

{Given} that the dust reverberation mapping technique {may} relate AGN V magnitudes and distances \citep{10.1088/2041-8205/784/1/L11},   we {additionally} mapped the  expected detection distance across V magnitudes (see Figure  \ref{fig:gridgravity} (d)). {This is accomplished by} solving the simple equation for `maximum detection distance' \citep{10.1007/BF00644218}, {which has been adjusted}  for our purposes. As the astrometric signature decreases with increasing distance, and the measurement error increases as the object (with absolute magnitude  $M$) becomes fainter with increasing distance, the maximum detection distance is the solution for $D$ of the equation 
\begin{equation}
\tilde{\mathcal{A}}=3\times \sqrt{\frac{2}{6}}\times(\sigma D\times10^{\frac{1}{5}(M-15)}),
\end{equation}
where {the right side of the equation represents}  the three times the error 
in one year normal point \citep[see][]{10.1007/BF00644218}
assuming a single observation error of $\sigma\sim 6 \mu$as for objects with $V=15$, and factor$\sqrt{2/6}$ for converting  single-point measurement error (in two coordinates) to one year normal point error if GRAVITY+   made six observations per object in {a} year.

{In the following section,} we parameterise our simulations across a grid of \eCBSMBH\, orbital parameters,{ and display} the results over a grid of the most important parameters {impacting}  $\mathcal{S}$: period, total mass, and eccentricity.

\section{Synthetic survey data and orbit fitting results}\label{sec:results}

\subsection{Multi-data simulation}

We simulate astrometric and RV data to evaluate the detectability of a \eCBSMBH \, {when its orbit is incomplete.}
{For binary stars, Aitken’s criterion typically needs $f_{\rm orb} \gtrapprox 0.75$, where $f_{\rm orb}$ is the portion of the observed orbit \citep{Ai64,10.1051/0004-6361/201322649}. 
        In this section, we analyse  incomplete orbit measurements of $f_{\rm orb}\sim 0.05-0.11$,}  in which a binarity signal is barely detectable because of a limited number of observations, which may be realistic for some \eCBSMBH s . {The fitting} procedure on an incomplete data set {might} result in multi-modal MCMC posterior distributions \citep[as confirmed in exoplanet detection][]{10.1086/500802}.

{Here, we let} $\mathcal{W}$ {be} a  space {composed of} vectors containing the \eCBSMBH\,  parameters $\mathbf{w_k}=(M_{k}, ,\tilde{a}_{k},e_{k},i_{k},P_{k},\Omega_{k},\omega)_{k}\in \mathcal{W}$, {where} $k=1,2$. {Given these vectors, the binary is `observed’ at times 
:
        \begin{equation} 
        t_{n}=f_{\rm orb} P \frac{k-1}{N-1},
        \end{equation}
        \noindent for $k=1,..,N$ so that $f_{\rm orb}$ is uniformly sampled. At each time  $t_n$, the `observed' multi-data set  is obtained as:} $\mathcal{M}(\mathbf{w_{k}},t_{n})$.
{The Bayesian approach does not require uniform sampling, and therefore it is assumed here for simplicity. The obvious alternative is random sampling, which might be  an unrealistic model for GRAVITY+ observational time baselines. }

The errors for each artificial observation are independent and identically distributed, resembling white noise at the level of $5\%$\footnote{\citet{10.3847/1538-4357/abc24f} generate mock astrometric data, adopting errors of $4 \mu$as in both astrometric coordinates based on current GRAVITY parameters, which are about $13\%$ and $26\%$ of the largest  astrometric offsets  that these authors estimated for SDSS J140251.19+263117.5}. 
{In order to avoid using} the same model for  the observations and finding the inverse solution \citep[see][]{link.springer.com/book/10.1007/b138659, 10.1051/0004-6361:200810288}, additional jitter was {added} in the model when {simulating} the data.  
Otherwise, the simulated observations and the corresponding solutions would only {aid} in {examination of} the model, which is not {always} encountered in reality \citep{10.1051/0004-6361:200810288}.

{
        In addition, three models of NIR} continuum emission photo-centres ($ \mathbf{r_{m}}$) are included in the synthetic observations:

\begin{equation}
\mathbf{r_{m}} \propto
\left\{
\begin{array}{ll}
\mathbf{c}={\rm const}&   \\ \\
\displaystyle 
\frac{\int^{b}_{a} \mathbf{r}_{\rm sub}\,g(\mathbf{r}_{\rm sub})dL}{ \int^{b}_{a} \,g(\mathbf{r}_{\rm sub}){dL }  }, & a\leq L\leq b, \\  \\
\displaystyle \overline{\mathbf{r}}_\mathrm{sub}\cdot \sqrt{ (1+\mathrm{sinc}\frac{2\pi t_{d}}{P}\sin{\frac{2\pi t}{P}})}

&\mbox{if } t_{d}=\frac{\|\overline{\mathbf{r}}_{\rm sub}(t)\|}{c}
\end{array}
\right.
\label{rsub}
\end{equation}

\noindent where $L$ is the length of the {arc  determined} by the intersection of the sublimation radius {bound} to the secondary SMBH ($\mathbf{r}_{\rm sub}$) and  CBD ($\mathbf{r}_{\rm CBD}$). {For simplicity, the} density of the sublimation surface arc between any two loci $g(\mathbf{r}_{\rm sub})$ is {considered} to be {one}. 
{During }\eCBSMBH\, orbital motion, {the} intersection points of the CBD and the sublimation ring are {determined} for each time instance $t$  (see the left panel in Figure \ref{model1}). The average dust ring offset is {assumed to be}
$\|\overline{\mathbf{r}}_{\rm sub}\| \sim 150 $ld.

\begin{figure*}[ht!]
        \includegraphics[ width=0.49\textwidth]{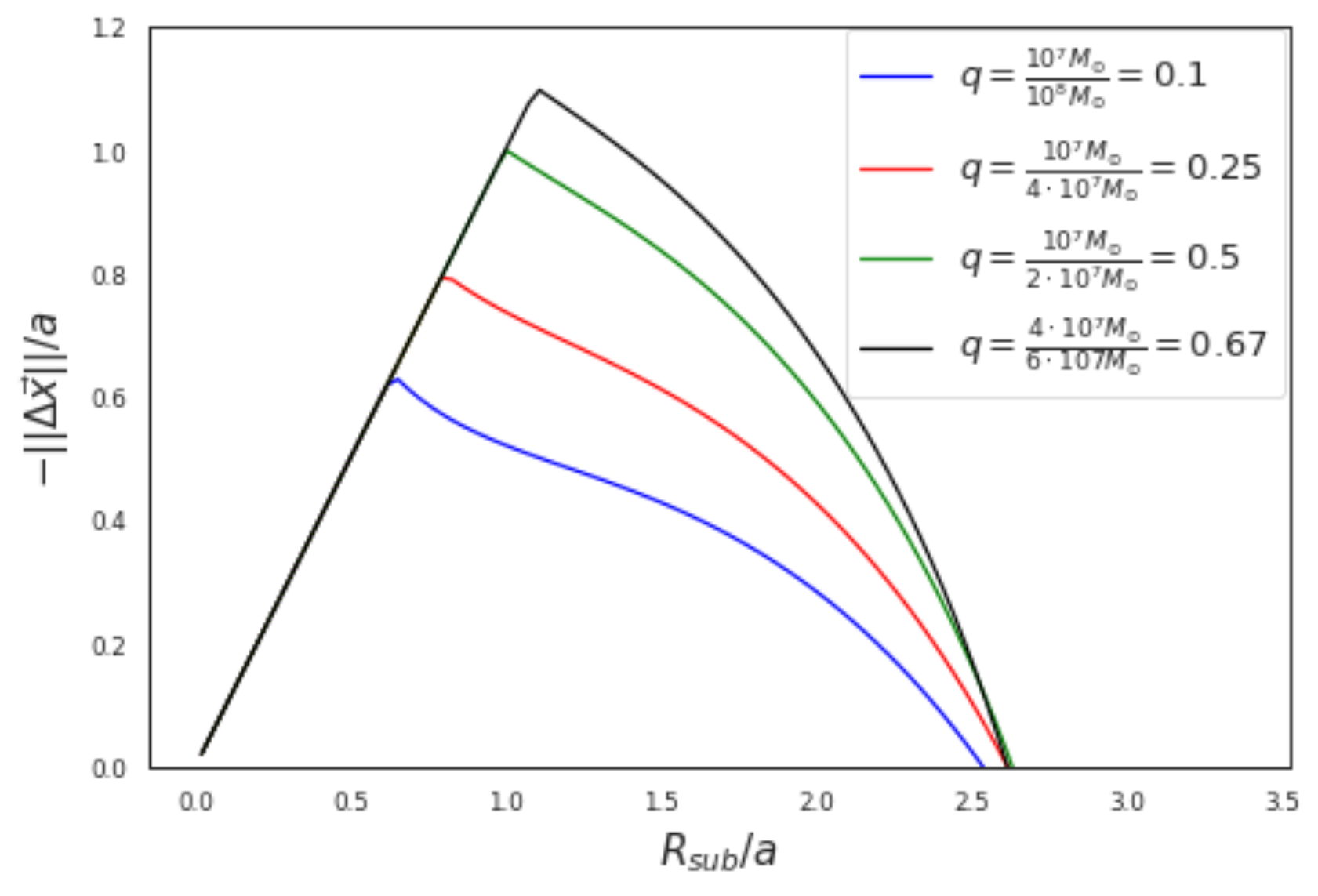}
        \includegraphics[ width=0.49\textwidth]{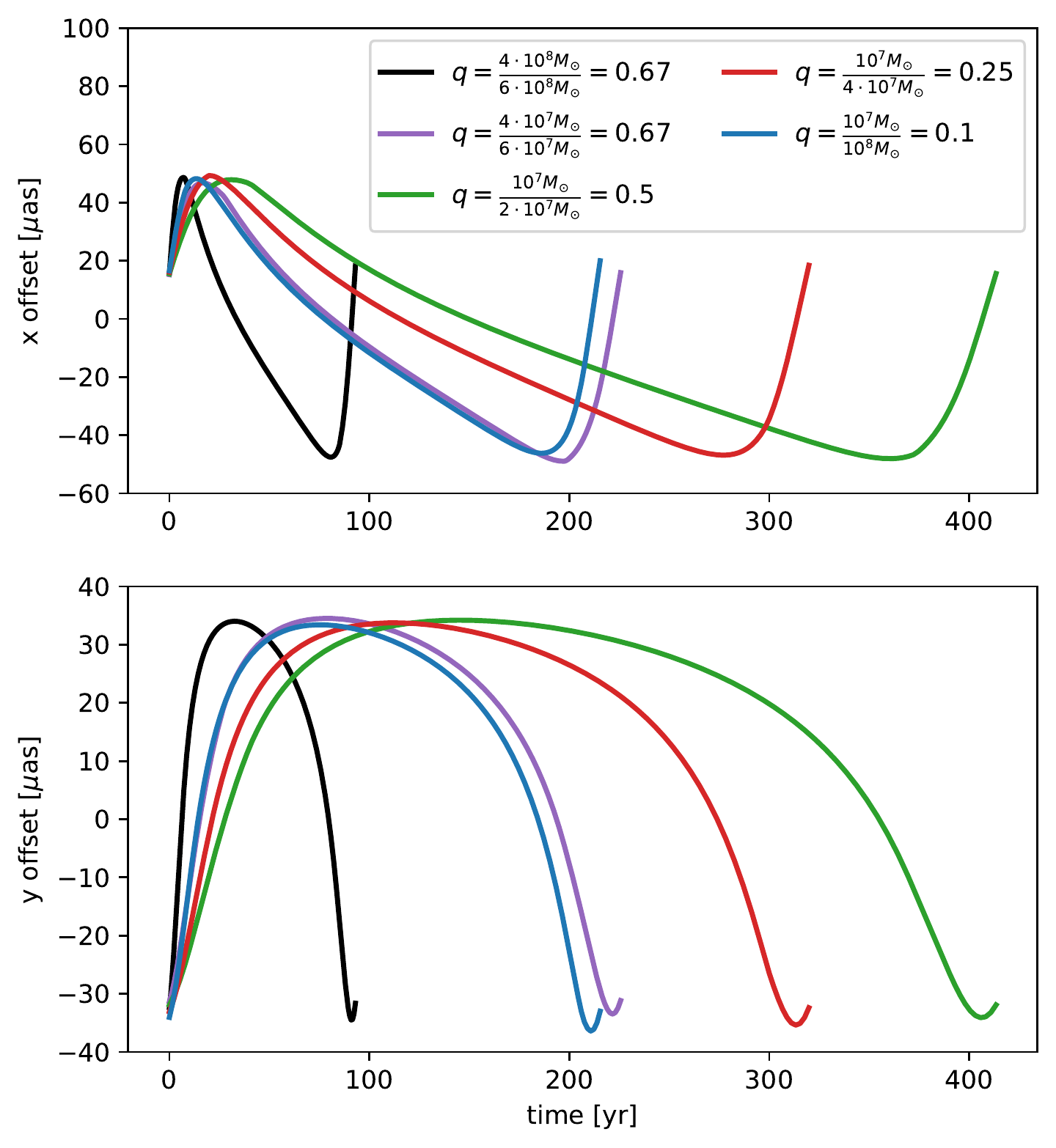}
        \caption{{The temporal evolution of the offset in both astrometric coordinates across one orbital period for different \eCBSMBH parameters (see the text).}
                \textit{Left} 
                General evolution of the astrometric offset of the photo-centre relative to the secondary
                with respect to $\Rsub/a$ at half orbital period
                (see the second branch in Equation \ref{rsub} and Appendices A-B)  for {\eCBSMBH}  of {varying  mass ratio} ($q$) and at a  distance {of} 700 Mpc.   \textit{Right} Evolution of the photo-centre  offset in two astrometric coordinates with respect to{ varying  mass ratio} and complete orbital period. The mean distance between the components is 100ld, and the assumed distance {between} the object and {the observer}  is 700 Mpc.  }
        \label{model}
\end{figure*}

If the continuum emission is stationary, that is, {fixed to } the inner edge of the CBD ($\mathbf{r_{m}} \propto \mathbf{c}$), then its position with {regard} to the  \eCBSMBH\,  barycentre is constant  (see the first branch of the Equation \ref{rsub}). However, the centroid of the arc of the dust-ring {split} by the CBD,\,{as seen} in the second branch,  will be {the location of} 
the evolving continuum emission {photo-centre}.
Deriving its analytic form {is simple}  (e.g.  see the Appendix A) and can{ take several} forms {depending on} the coordinate system.
In general, {the behaviour of the astrometric offset of the photo-centre relative to the secondary for the non-static} continuum emission of the \eCBSMBH\,(see the left panel in  Figure \ref{model}) follows the {trend} found in a circular CB-SMBH  \citep{10.3847/1538-4357/abc24f},  with {slight modifications} due to eccentric motion.
{We find that} anti-aligment of the angular momenta of the sublimation surface ($\mathbf{L}_{\rm sub}$) and CBD ($\mathrm{L}_{\rm CBD}$) {has no effect on the overall behaviour} of the photo-centre of the non-static hot-dust emission seen in   Figure \ref{model}.
We show the temporal evolution of the offset in both astrometric coordinates across one orbital period for \eCBSMBH s\, of various masses  $\Big\{(M_{1}, M_{2})\Big\}=\Big\{(60,40),(6,4),(2,1),(4,1),(10,1)\Big\}
\times 10^{7}M_{\odot}$,  fixed orbital parameters $e_{1}=e_{2}=0.5,\Omega_{1}=250^{\circ},\omega_{1}=220^{\circ}, \Omega_{2}=70^{\circ},\omega_{2}=40^{\circ}, \Rsub=150 \mathrm{ld,}$ and a mean mutual distance of 100 ld for a non-static continuum emission model  (see Figure \ref{model} right panel).
{Finally}, the sinusoid variation of the continuum emission photo-centre in the dust torus {is represented} by the third branch of Equation  \ref{rsub}.

{The simulated observational campaigns are constructed by  $\mathcal{C}=(N, \mathcal{T},P)$, each  with a different total number of observations (N), monitoring campaign length ($\mathcal{T}$), and  \eCBSMBH\, orbital period ($P$). }
When simulating the measurements, the monitoring campaign parameters are set to 
$\mathcal{C}_{1}=(10,12 \mathrm{yr}, 221 \mathrm{yr})$ and  $\mathcal{C}_{2}=(14,10 \mathrm{yr}, 93.75 \mathrm{yr})$. In these scenarios, the values of all the other  orbital parameters, including the masses  and coplanarity  of the \eCBSMBH\, and  CBD  were fixed.

\subsection{Orbit fitting}

{Historically, the incompleteness of binary orbits has been handled by scanning parameter space for the global minimum, which may be the closest practical approximation to the truth, or by establishing a complete set of acceptable orbits and then computing an average \citep[e.g. see][and references therein]{10.1051/0004-6361/201322649}.}
{However, the} posterior probability distribution of model parameters contains all of the information in a Bayesian framework.
{By scanning parameter space, the posterior means of the orbital elements or any function of them can be determined without finding minima.}

\begin{table}
        \caption{Priors for the model of the motion of the secondary component.
                The mean value and standard deviation of the \eCBSMBH\, parameters, as stated in the text, {enter} the normal distribution, {whereas} the physically {permissible} interval of the \eCBSMBH\, parameter {determines} the uniform distribution.}
        \centering
        \begin{tabular}{lc}
                \hline\hline
                Parameter & Distribution \\
                \hline
                P[yr] & $\log (P/({\rm yr}))$ is Normal (2.31, 0.5)  \\
        $M[M_{\odot}]$ &$\log (M/M_{\odot})$ is Normal (7.9, 0.05) \\
        e & Uniform(0.,0.7) \\
        $\Omega[{\rm rad}]$ & Uniform(0,$2\pi$)  \\
        $\omega[{\rm rad}]$ & Uniform(0,$2\pi$)  \\
                        \hline
        \end{tabular}
        \tablefoot{It is expected that \eCBSMBH\, will converge to a critical eccentricity value $0.55 < e_{c}< 0.79$ if the ratio of the location of the inner rim of the CBD with respect to the barycentre of the binary is between 2 and 2.5 \citep{10.3847/0004-637X/817/1/70}.
        }
        \label{priors}
\end{table}

As Bayes' theorem indicates, by combining two or more measurement methods {(e.g.  astrometry and radial velocity in our case)}, we can infer more information about the observed target than {relaying on} a single method:

\begin{equation}
\mathcal{P}(\mathrm{param}|{\mathrm{data}})\propto \mathcal{P}(\mathrm{data}|\mathrm{param})\times\mathcal{P}(\mathrm{param}),
\label{par}
\end{equation}
where $\mathcal{P}(\mathrm{param}|\mathrm{data})$ is the posterior distribution, {which provides} the probability distribution of the full Keplerian model parameters given the observed data (i.e. $\mathbf{\Xi}^{\rm o}(t),\dot{z}^{\rm o}(t) $); $\mathcal{P}(\mathrm{param})$ is the prior distribution, {which reflects the prior}  belief  about the values that the unknown parameters $\mathbf{w}$ can take before observations are obtained; and 
$\mathcal{P}(\mathrm{data}|\mathrm{param})$ is the likelihood distribution, which gives the probability distribution of {data} values  that can be measured for the given {parameter} values.
{Because} the astrometric data and radial velocity are measured independently, Equation \ref{par} {may be} rewritten {as follows}:
\begin{equation}
\mathcal{P}(\mathbf{w}|\mathbf{\Xi}^{\rm o}(t),\dot{z}^{\rm o}(t) )\propto\frac{\mathcal{P}(\mathbf{\Xi}^{\rm o}(t)|\mathbf{w})\mathcal{P}(\dot{z}^{\rm o}(t)|\mathbf{w}))}{\mathcal{P}(\mathbf{\Xi}^{\rm o}(t)) \mathcal{P}(\dot{z}^{\rm o}(t))}\mathcal{P}(\mathbf{w}).
\label{eq:par1}
\end{equation}

In the Bayesian formulation,  an increase in information is reflected either as an increasing set of model parameters or narrow parameter densities. 

For all parameter combinations, the posterior probability distribution is calculated by integrating Equation  \ref{eq:par1}. 
{However}, the parameter space $\mathcal{W}$ (defined in Sec. \ref{sec:astr})   is large because of the high dimensionality of the Keplerian model.
To estimate the posterior distribution in {an  acceptable period} of time,  we {used the }numerical sampler to efficiently explore the parameter space $\mathcal{W}$. \texttt{PyMC3} \citep[Python package for Bayesian inference][]{10.7717/peerj-cs.55} is adequately sampling the posterior without exploring {unfeasible parts of parameter space}. 
We also summarise the calculated posterior distribution  {by the region with} the highest posterior density (HPD),\, {and the lowest} volume  of all 
$(1-\alpha)\%$ credible regions $\mathcal{C}_{\alpha}$,  {so that the following holds}:
\begin{equation}
\int_{\mathcal{C}_{\alpha}} \mathcal{P}(\mathbf{w}|\mathbf{\Xi}^{\rm o}(t),\dot{z}^{\rm o}(t) ) \,d\,\mathbf{\Xi}^{\rm o}(t)\,d\dot{z}^{\rm o}(t)\geq 1-\alpha 
\label{eq:credible}
.\end{equation}
For the unimodal posterior, the HPD region {consists} of a single region of the parameter space. However, if the posterior is multimodal, {the HPD may consist of}  an ensemble of disjointed regions, the estimate {of which} is {typically} more computationally {expensive}.
{The HPD corresponds to locating the true parameter in the smallest possible region of the sample space with a given probability $(1-\alpha)$.}

{The use of} Bayesian inference between RV and astrometric data {allows} the model parameters to  {be fit to} the artificial data containing three types of perturbations. 
{The \texttt{PyMC3} NUTS sampler is an MCMC technique that avoids random walk behaviour and enables faster convergence to a target distribution. This has the advantage of not only being faster, but also allowing complex models to be fitted. Two chains of the \texttt{PyMC3} NUTS sampler were run. The beginning state ${w(0)}$ of each
chain is picked at random from the prior distribution, affecting only the pace of convergence.
        We had 5000 samples per chain to auto-tune the sampling algorithm and 4000 productive draws  yielding a total of 20000 samples per chain. It is worth noting that, in addition to parameter priors, the model considers observed data while constructing the posterior distribution.}

{
        For the purpose of this study, we devised the following protocol.  Three groups of tasks are identified: (1) the simulator  generates the simulated observations, assuming specific characteristics of the {\eCBSMBH;} (2) the solver uses the simulated data to find `solutions' of {\eCBSMBH} orbital parameters; and (3) the evaluation takes both the `truth', that is, the input parameters, from the simulator and the solutions from the solver, compares the two, and draws a set of conclusions.
        All tasks require a separate set of simulations, and they are carried out in several steps:}
{
        \begin{itemize}
                { \item Simulation of the observation: (a) For the synthetic observational campaign  $\mathcal{C}_{1}=(10,12, 221\mathrm{yr})$, we assume that the binary is observed with  the following parameters:  
                        $M_{1}=6\times 10^{7} M_{\odot},\, M_{2}=4\times10^{7} M_{\odot}, \,e=0.5, \Omega_{1}=\Omega_{2}=180^{\circ},  \omega_{1}=181^{\circ},  \omega_{2}=1^{\circ}$,   $\Rsub=150\, \mathrm{ld}$, mean mutual distance of 100 ld, an object distance of $\sim 700$ Mpc, and  an observer position angle of $\pi/6$. The average distance of the best ten GRAVITY+ circular binary candidates is 700 Mpc.
                        (b) For comparison, we consider the    monitoring campaign parameters  $\mathcal{C}_{2}=(14,10,93.75 \mathrm{yr})$,  {\eCBSMBH} parameters of  $M_{1}=6\times 10^{8} M_{\odot}$, \, $M_{2}=4\times 10^{8} M_{\odot}$, and the same remaining orbital parameters as in (a)
                        .}
                {\item Solver: The prior probability distributions for the model parameters that are {assumed to be independent} are shown in Table \ref{priors}.
                        {Physical and geometric considerations lead to natural choices for the prior PDFs for most of the model parameters.}
                        We choose normal priors on ${M}$ centred on $10^{7.7} M_{\odot}$  {as  estimates of SMBH mass functions peak between $10^{7}  M_{\odot}$ and $10^{8}  M_{\odot}$ for quasars at $z\sim 0.5-1$ \citep{10.1088/0004-637X/692/2/1388}} .
                        The normal priors on $P$  are centred on  $200$ yr, {because the period for the binary at a mutual distance of 0.05 pc and total mass of  $10^{7.7} M_{\odot}  $  would be $\sim 150$ yr. }
                        {We adopt non-informative,} uniform priors on orbital angles. {The bounds of the uniform} PDF {} are defined {in such} a {way} that the tool does not explore unphysical domains.
                        Because of the uncertainties in the   artificial data,  the likelihood distribution of the fitting procedure is chosen as a Gaussian distribution  centred at both astrometric variables $\Xi(t)$ and radial velocity $\dot{z}$
                        given by  Equation \ref{eq:compmodel} with  standard deviations of $5\%$. The error priors   are drawn from normal distributions {that are} centred at  expected errors of artificial data {and have a} standard deviation of $5\%$.}
                { \item Evaluation:  
                        It would be {useful to assess} how the detection algorithm performs {across the entire parameter set}. However, due to the great complexity of the problem, we use $(M,P,,e)$ as proxies to understand the behaviour of the {\eCBSMBH} orbital solutions. We compare the results  obtained from fitting observations from two different campaigns in this section and evaluate additional considerations in subsequent sections.}
\end{itemize}}

\begin{figure*}[ht!]
        \includegraphics[ width=0.45\textwidth]{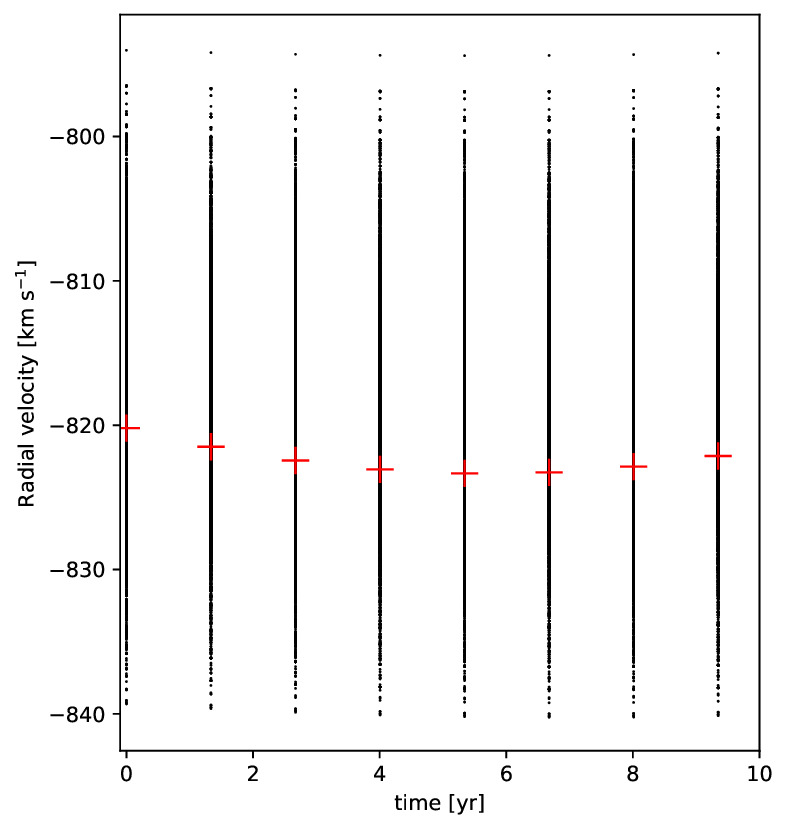}
        \includegraphics[ width=0.45\textwidth]{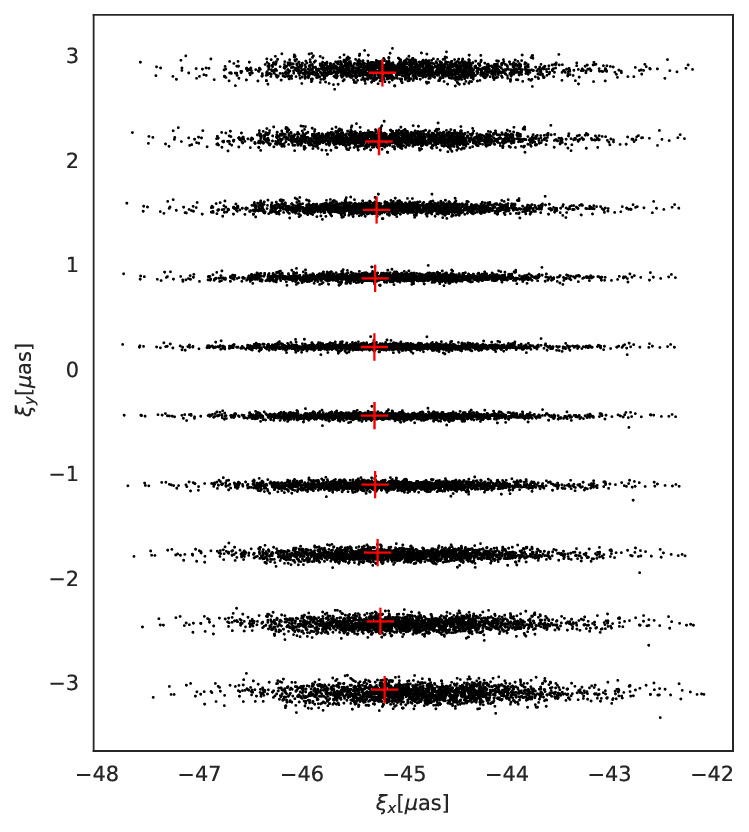}
	
        \includegraphics[ width=0.45\textwidth]{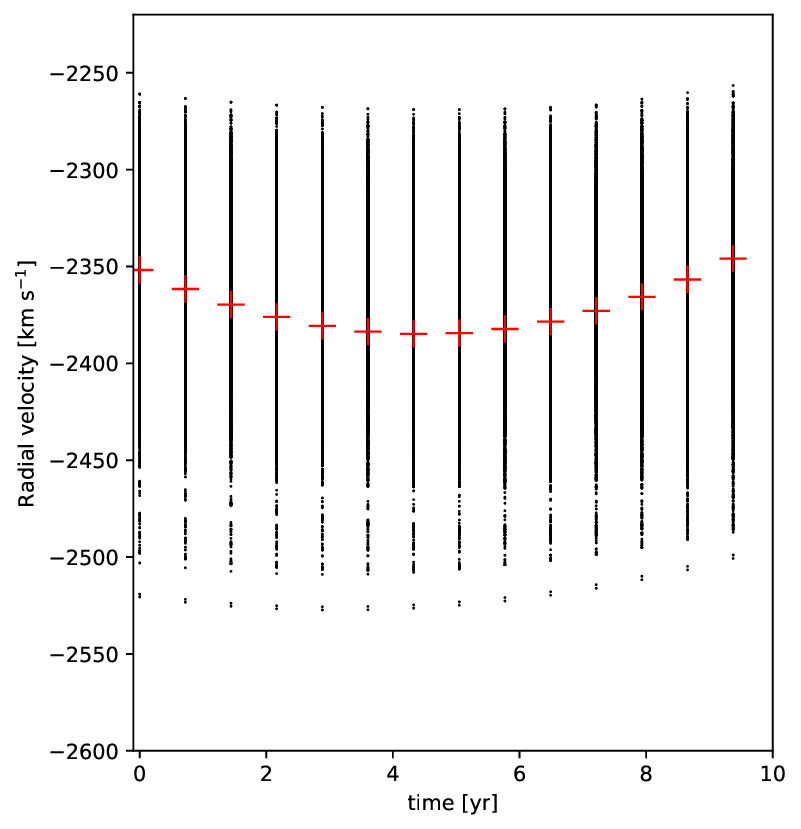}
        \includegraphics[ width=0.45\textwidth]{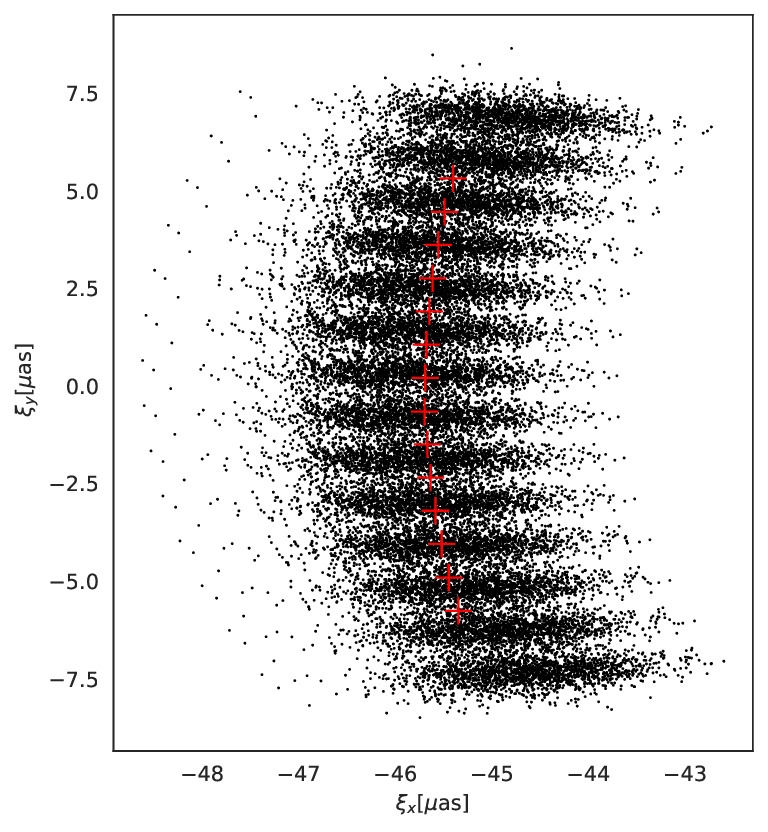}
        \caption{Simulated incomplete observations of RV  and  astrometric data of the secondary SMBH (red crosses) and the distribution of the modelled posteriors (black dots) for {the evolving model} given in Figure \ref{pymc3} ($f_{\rm orb}\sim 5\%) $, upper row) and Figure \ref{pymc31} ($f_{\rm orb}\sim 10\%) $, {bottom} row). 
                \textit{Left column} RV data for the secondary SMBH and {the} posterior distribution of simulations over time. {We note here the RV curve gradient  differences: almost constant (upper row) and variable (bottom row)}. \textit{Right column} Astrometric observations of the secondary SMBH  and {the} posterior distribution of simulations over time.   
        }
        \label{predict}
\end{figure*}

\begin{figure*}[ht!]
        \includegraphics[ width=0.8\textwidth]{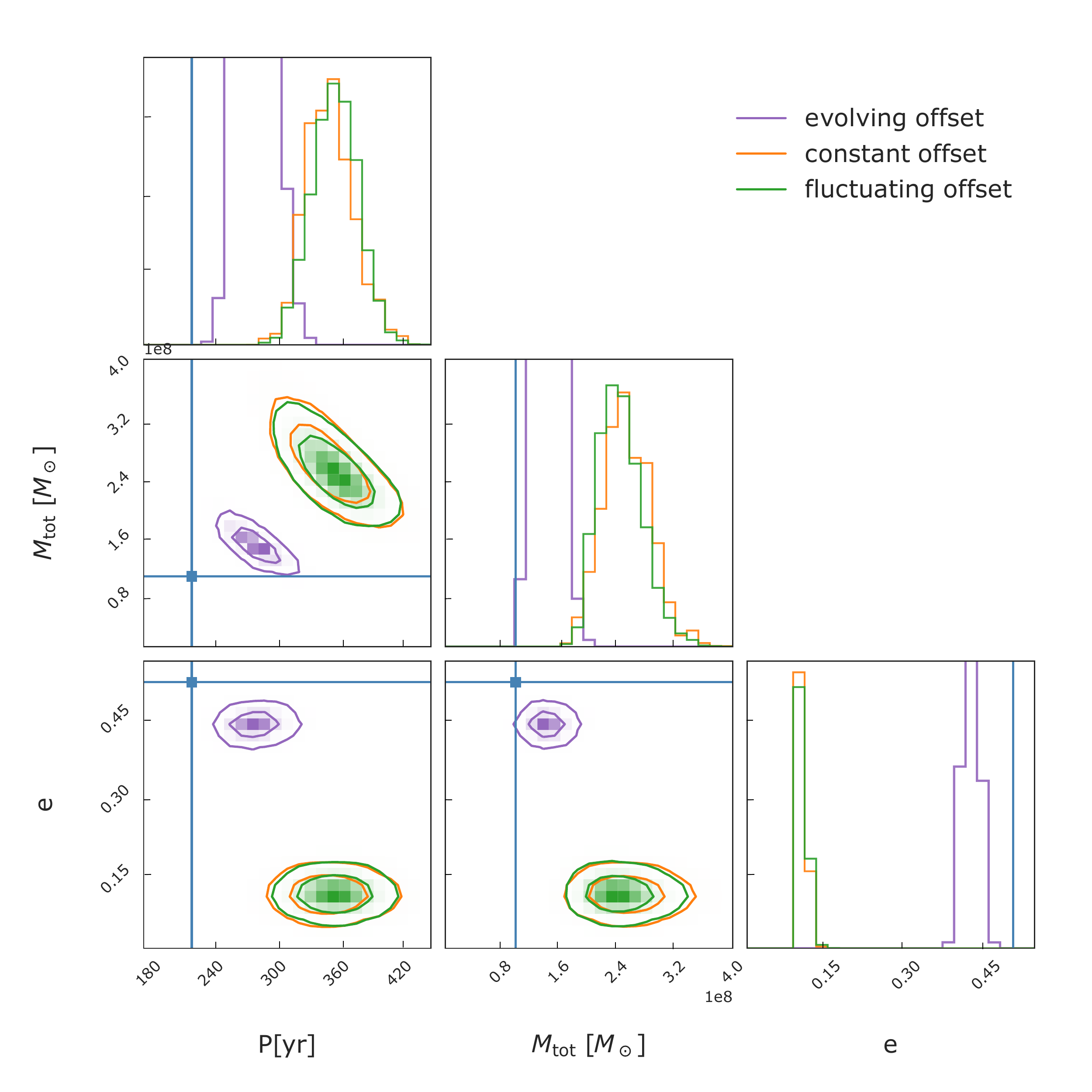}
        \caption{Joint posterior distribution resulting from the Bayesian inference for three different perturbations in simulated data of \eCBSMBH\, with monitoring campaign $\mathcal{C}_{1}=(10,12 \mathrm{yr}, 221 \mathrm{yr})$, binary components  
                $M_{1}=6\times 10^{7} M_{\odot}, M_{2}=4\times10^{7} M_{\odot}$, and a mutual mean distance of 100ld (other orbital parameters are given in the text) for (i) a constant offset hot-dust continuum, (ii) a fluctuating hot-dust continuum offset, and (iii) the evolving continuum model where the offset tracks the  orbital motion of the secondary. The diagonal plots show the marginal distribution for each parameter (the projection of the posterior onto that parameter). The  contour plots represent the  95$\%$ and 68 $\%$ credible regions {(not proportional to one- and two-sigma level)} described in Equation \ref{eq:credible}. The solid blue lines represent the true parameter values.
                The covariance between the total mass and period seen here arises from  Kepler’s third law implemented in our model.
        }
        \label{pymc3}
\end{figure*}

{
        As an example, Figure \ref{predict}  {shows} the simulated RV and astrometric data for evolving offset obtained from the simulator for campaigns $\mathcal{C}_{i}, i=1,2$.
        The solver performed the Bayesian fitting procedure to determine how well the orbital parameters of an {\eCBSMBH} can be measured for two different campaigns determined by the simulator. The distributions of  the modelled posteriors are depicted in  Figure \ref{predict}.}
{Figure \ref{pymc3} shows} the corresponding densities of orbital parameters \footnote{The contour levels are not 1 and 2 sigma levels (which in two dimensions correspond to $39\%$ and $86\%$ contour levels). As posterior distributions are not perfect Gaussians  in general, there are no preferences in choosing one or other definition.}  ($P, M$ and $e$) for campaign $\mathcal{C}_{1}=(10,12, 221\mathrm{yr})$.  We {can observe that} the maximum a posteriori estimates of these densities are {fairly} close to the original binary parameters. The period, eccentricity, and total mass are {all feasible, although with a reduced degree of certainty.}

\begin{figure*}[ht!]
        \includegraphics[ width=0.8\textwidth]{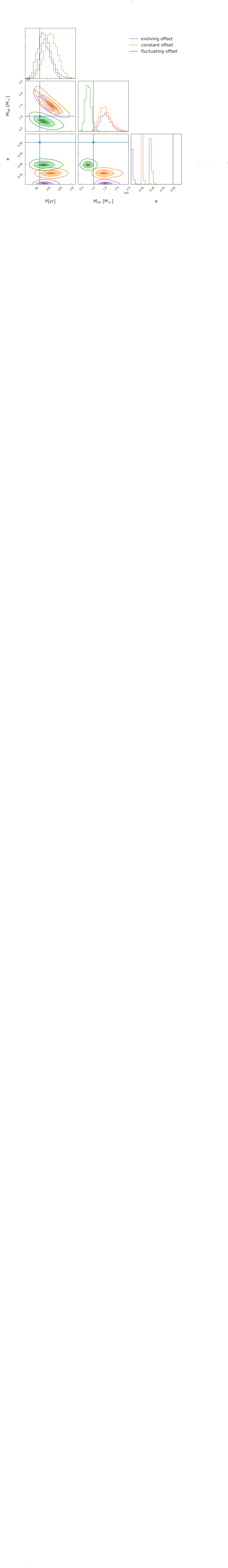}
        \caption{Same as Figure \ref{pymc3} but for an \eCBSMBH \,with monitoring campaign $\mathcal{C}_{2}=(14,10 \mathrm{yr}, 93.75 \mathrm{yr})$, binary components, $M_{1}=6\cdot10^{8} M_{\odot}$, $M_{2}=4\cdot10^{8} M_{\odot}$, and a mean mutual distance of 119 ld.}
        \label{pymc31}
\end{figure*}

{For the second type of monitoring campaign   $\mathcal{C}_{2}=(14,10,93.75 \mathrm{yr})$, the simulated observed data  {span} $\sim 10\%$ of the orbital period (see Figure \ref{pymc31} and \ref{predict}). The solver found} that  the mass, orbital period, and eccentricity are more likely to be {reconstructed when using a data set based on} a model with evolving dust 
constant and variable dust offset models. 
{In contrast}, for objects of {greater} mass, the inferred periods are closer to the real value (compare results in Table \ref{poster} vs. those in  Table \ref{poster1}), 
{{as well as} {the posterior distribution of}  RV curves and astrometric orbits for {fitted} parameters }(see Figure \ref{predict}). 

\section {Refining the  {\eCBSMBH} detectability}\label{sec:discussion}

Motivated by the {upcoming}  GRAVITY+ instrument operations, we {evaluate} the detectability of \eCBSMBH\, systems {using} simulated multi-data {sets} (astrometric and RV). We extended the investigation by \citet{10.3847/1538-4357/abc24f} to a broader parameter range ({particularly} \eCBSMBH\, eccentricity) while {accounting for} the realistic and unfavourable percentage of \eCBSMBH\, orbits covered by observations.
{For} the continuum hot dust emission, we use constant, evolving, and fluctuating models, {as well as the } \eCBSMBH\, dynamical model for the astrometric and RV data {simulations}.
We quantify \eCBSMBH\, detection {by the GRAVITY+ instrument} in terms of a simple detectability statistics  as well as  Bayesian inference of {an incomplete ($f_{\rm orb}\sim 0.1$)} \eCBSMBH\, orbit using hot-dust emission models.
{Based} on MCMC orbit fitting, we {find} that the evolving hot-dust emission model is more {resilient}\footnote{A resilient model will perform well on a wide range of $f_{\rm orb}$  below  optimal values. It will also perform better for longer orbital periods.} when  recovering the basic orbital parameters of the \eCBSMBH \, than constant and fluctuating models.

{Besides the above general outline, {\eCBSMBH} detection refinements based on the additional considerations, for example, variation of  $q, f_{\rm orb}$ parameters (Section \ref{sec:vari}),  
{ability to retrieve}  orbital eccentricity  from radial velocity and acceleration data (Section \ref{sec:refine}), and refinement of binary detectability in contrast to other CBD phenomena (Section \ref{sec:cbdrefin}),  are discussed below. We conclude this section by {recapitulating} the   SARM
technique, which can be employed  for  refinement of binary detection either through follow-up or as an independent binary detection tool (Section \ref{sec:sarm})}.

\subsection{Refinement of binary detectability based on variation of $q, f_{\rm orb}$ parameters}\label{sec:vari}

{When} formed in minor galactic mergers, it {appears} that typical \eCBSMBH s\, could have different mass ratios ($q$). If a binary is the outcome  of a major merger, then the mass ratio can be moderate and  deviate from unity $q\sim 0.5$ \citep{10.1086/497108}. 
Accounting for galaxy luminosity statistics {leads to the conclusion} that most galaxy interactions {feature} central black holes with  mass ratios in the range of  $\frac{1}{3}< q<\frac{1}{30}$
\citep{10.1088/1742-6596/122/1/012040}.

{
        Two binaries should have slightly different astrometric signatures if their mass ratios are slightly different.
        If we compare a   binary with parameters $q=0.1, M_{1}=10\cdot10^{6} M_{\odot}, M_{2}=10^{6} M_{\odot} $ to a  binary with $q=0.11, M_{1}=10^{8} M_{\odot}, M_{2}=11\cdot10^{6} M_{\odot}$, then  the latter system  will  have  an 11\%\ larger astrometric signature.
        }
{Consider the impact} of extreme  ratios of small integers (smaller or equal to 10), $q=1/10=0.1$, extreme ratios of large integers $q=67/100=0.67$, and  non-extreme but unequal mass ratios, $q=0.25,0.5,$   on the astrometric signal detectability and {astrometric data}.
The astrometric S/N and detection distance for  \eCBSMBH s\, with equal mass ratios are {greater} than those with slightly non-equal mass ratios. The best GRAVITY $+$ {circular} targets 
are  {distinguished} by {their high} S/N and large detection distance. Interestingly, the time evolutions of astrometric offsets are {clustered} into {two distinct groups} based on{ two types of SMBH mass ratios}: extreme
and {moderately} unequal (see Figure \ref{model} (b)).

{After describing the difference in} time evolution of astrometric {offsets caused by different mass ratios, let us now address the incompleteness of }orbits ($f_{\rm orb}\sim 0.05-0.1$),  when  any time instance of observation {meets the condition} $\frac{t_n}{P}\rightarrow 0$ (see e.g. Figure  \ref{predict}).
{A basic inspection of} Equation \ref{eq:compose} shows that in  such {small} time instances, vector components vary {little} and can correlate.{
        When assuming $e\approx 0,$ then the following expressions hold true: $E(t)\sim 2\pi n (t-t_{0})$ and
        $\mathbf{r}(\mathbf{w},t)\sim a_{\bullet} ( \mathbf{p}\cos(\omega+2\pi n t) +\mathbf{q} \sin(\omega+2\pi n t))$.
        We can expect small perturbations of the model
        $\mathbf{r}(\mathbf{w},t)-\mathbf{r}(\mathbf{w+\delta w},2\pi n (t-t_{0}+\delta t_{0}))\approx 0$ for small perturbations of the vector of parameters $\mathbf{w}$.
        However, this implies that $\delta w\sim n\delta t_{0}$ (we note that $n$ scales inversely with the period of the binary), resulting in a correlation between $\omega$ and $t_{0}$.}
{Furthermore,} the right-hand side of the Kepler equation will {converge} to {extremely} small {values}.   { These tiny effects can distort} posterior PDFs of parameters (see Figure \ref{pymc3}), { causing} orbital parameters to be underestimated or overestimated.
{A further challenge} is that three parameters in the model  $(P, e, \omega)$ {contribute} to the astrometric offsets and radial velocity in a non-linear {fashion}.
{Moreover,}  the values of the parameters {under discussion} typically differ {across} orders of magnitude.  {The binary} total  mass has magnitudes of order $10^{7-10}M_{\odot}$ {yet} periodicity {spans}  $10-10^{3}$ yr.  
Another issue is that posteriors of mass and eccentricity are often highly correlated, {leading to substantially} slower Markov chain convergence.

{Even for incomplete binary orbits  ($f_{\rm obs}\sim 0.05-0.1$), we see impacts of Bayesian inference (see Equation \ref{eq:par1}); for example, conjugated multiple observational techniques  generate more information on the system
– either in a narrower posterior parameter density (Figures \ref{pymc3} and \ref{pymc31})
{ or in the potential of include  additional parameters in the model
} {or in the capability to include additional parameters in the model.}} Figures \ref{pymc3} and \ref{pymc31} show {how} posterior PDFs differ from perfect Gaussian distributions, particularly in the case of eccentricity. However, the {vast majority} of prior PDF samples have been discarded, and only {a small} subset of periods, masses, and eccentricities are {compatible} with the data. Even {distorted} posterior PDFs can {give} a very  informative prior PDF for the design of future surveys  \citep{10.3847/1538-4357/aa5e50}.
Tables \ref{poster} and \ref{poster1} {compare} the median values recovered from  three models; comparing posteriors   to the true values. 
{Except for the other true parameters}, only the total mass for  the \eCBSMBH\,  with {a} 221 yr orbital period falls within the central $3 \sigma$ credible intervals of the recovered value for the evolving model (Table \ref{poster}).

{At the same time}, binary masses {derived} from the constant and fluctuating models are {less well specified}.
In  {contrast}, for the \eCBSMBH\, with a {shorter} orbital period (93.25 yr), the {true} values of period and total mass fall within the central $3 \sigma$ credible intervals of the recovered values (see {Table} \ref{poster1}), but the true value for eccentricity is {within} the $3 \sigma$ of the recovered values. {We cannot\,{rule out} the possibility that the apparent effect of a {specific type} of `uncertainty principle' in  {determining} $P$ and $e$   {is} caused by {their different roles}: $e $ ({along} with $a$) accounts for the two degrees of freedom in the shape of the orbit, {whereas} orbital period $P$ locates {a} given object on its orbit at a given time.
        
        {Furthermore}, such uncertainty can {develop} as a result of a lack of knowledge about the true eccentricity distribution expected for \eCBSMBH s. Namely, we do not know whether  \eCBSMBH s\, can be {separated} into different subpopulations {based on}  eccentricity and  total mass as  is the case {with}  close stellar binaries \citep{10.1051/0004-6361:20021507}.  We {also} allow for an overall jitter in the radial velocity curve and astrometry data {to accommodate} for {imprecise} knowledge of data uncertainties and any intrinsic scatter. However, in model fitting, we did not consider jitter to be a non-linear parameter \citep{10.3847/1538-4357/aa5e50}. Finally, {the} Keplerian model is {dependent} on the data {rather than being} a simple function of the non-linear fitting parameters. 
        Increasing the non-linear\footnote{The linear parameters are algebraic combinations of $K$ and $\omega$, while $P, T_{0}$, and $e$ are non-linear parameters.} parameter $e$, for example,  has an effect on the model, not just because it is more eccentric, but also because the linear parameters have different values at this new $e$ value  \citep[see e.g.][]{10.1088/0067-0049/182/1/205}. A possible {approach  for these issues} would be to introduce fitting on analytically transformed orbital elements.  } 

\begin{table}
        \caption{True values to generate data and summary statistics from the posterior distribution  for three different models in Figure \ref{pymc3}. Columns E, C, and F {represent}  recovered values from evolving, constant, and fluctuating models, {respectively}. One-sigma errors indicating uncertainty are shown.}
        \centering
        \begin{tabular}{lcccc}
                \hline\hline
                {Parameter} & {True value}&{E}& {C} &{F}\\ 
                \hline
        P[yr] & 221&$282^{+15.6}_{-15.6}$ & $352^{+23.8}_{-21.9}$ &$358^{+21}_{-22}$  \\
$M[10^{8}M_{\odot}]$ &1 &$1.43^{+0.2}_{-0.2}$&$2.52^{+0.43}_{-0.3}$ & $2.47^{+0.33}_{-0.27}$ \\
e & 0.5&$0.42^{+0.01}_{-0.01}$ &$0.1^{+0.01}_{-0.002}$ & $0.1^{+0.001}_{-0.003}$\\
\hline
        \end{tabular}
        \label{poster}
\end{table}

\begin{table}
        \caption{\label{poster1}The column descriptions are {identical} to those  in Table \ref{poster}, but for the three  models {shown} in Figure \ref{pymc31}. }
        \centering
        \begin{tabular}{lcccc}
                \hline\hline
                {Parameter} & {True value}&{E}& {C} &{F}\\ 
                \hline
                P[yr] &93.75 &$100^{+7.4}_{-6.6}$ & $108^{+8.9}_{-8.7}$ &$99.5^{+9.04}_{-7.98}$  \\
        $M[10^{9}M_{\odot}]$ &1 &$1.51^{+0.24}_{-0.2}$&$1.48^{+0.28}_{-0.22}$ & $0.77^{+0.12}_{-0.12}$ \\
        e &0.5 &$0.3^{+0.01}_{-0.003}$ &$0.35^{+0.003}_{-0.001}$ & $0.39^{+0.005}_{-0.002}$ \\
                        \hline
        \end{tabular}
\end{table}

\subsection{Refinement of detecting binary  orbital eccentricity  from radial velocity and acceleration data}\label{sec:refine}

\begin{figure}[ht!]
        \includegraphics[ width=0.5\textwidth]{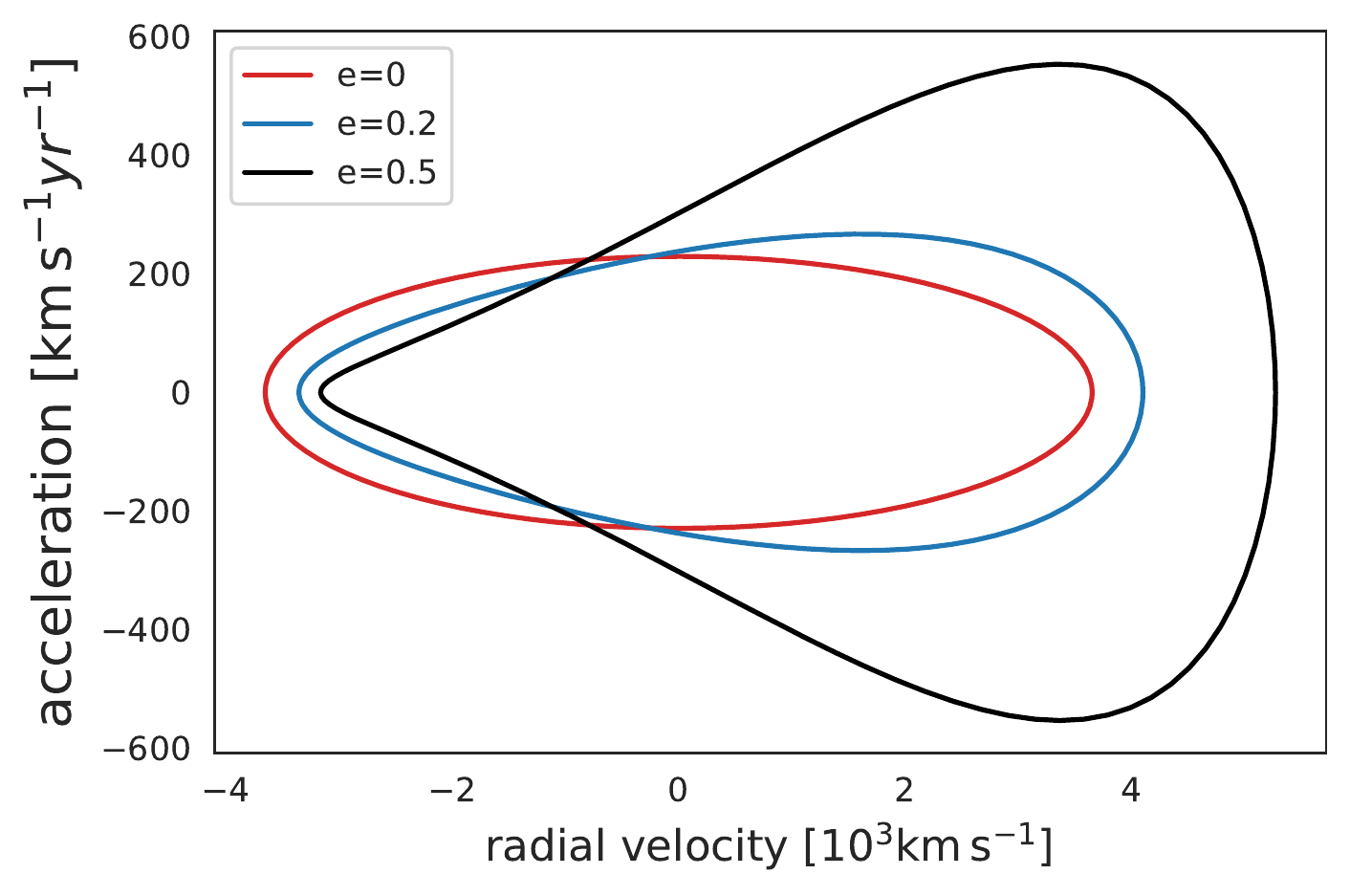}
        \caption{{Effects of  eccentricity on  deformation} of the curve {defined by} the radial acceleration of the secondary SMBH    and its radial velocity  relative to the primary.  \eCBSMBH\, parameters are  $P=100\, \mathrm{yr}, \omega=\pi/3, i=\pi/4, \tilde{a}=100 \mathrm{ld},$ and $ M_{1}+M_{2}=5.2\cdot10^{8}M\odot$.  }
        \label{vradvsaccel}
\end{figure}

{Another refinement} that has {yet} to be {addressed} is how to independently test the eccentricity of the \eCBSMBH\, orbit. We recall that {the expression} for relative radial velocity can be {provided} by

\begin{eqnarray}
V^{\rm r}_{\rm rad}=\dot{z}(t)&=&\frac{2\pi \tilde{a}\sin i}{P\sqrt{1-e^{2}}}(\cos (f+\omega)+e\cos \omega),
\label{radvel222}
\end{eqnarray}
\noindent where $f$ is {the} true anomaly, and $\frac{2\pi}{P}\tilde{a}=\sqrt{\frac{G(M_{1}+M{2})}{\tilde{a}}}$.
{It should be noted that the} radial velocity of the secondary with respect to the barycentre {is simply given by} $V_{\rm rad}=\frac{M_{1}}{M_{1}+M_{2}} V^{\rm r}_{\rm rad}$. 
{For}  moderate values of mass ratio and separations, the barycentre  will {be} outside the event horizon of the components \citep{
        10.1093/mnrasl/slv076}.
{ In such a case, the fluctuation} of the barycentric radial velocity of the  secondary can be {represented} as
\begin{eqnarray}
\frac{d V_{\rm rad}}{dt}=\ddot{z}(t)&=&\frac{-2\pi  \kappa}{P(1-e^{2})^{\frac{3}{2}}}\sin (f+\omega)(1+e\cos f)^{2},
\label{accel}
\end{eqnarray}
where 
$$
\kappa=\frac{2\pi \frac{M_{1}}{M_{1}+M_{2}} a\sin i}{P\sqrt{1-e^{2}}}.$$  
We can substitute $f$ in equation \ref{accel} as follows:
\begin{eqnarray}
\ddot{z}(t)&=&\frac{-2\pi  \kappa}{P(1-e^{2})^{\frac{3}{2}}}\sin \beta\cdot\left[1+e\cos (-\omega +\beta)\right]^{2},
\label{accel2}
\end{eqnarray}
where $\cos \beta=(\frac{\dot{z}}{\kappa}-e\cos\omega)$. 
The relation $\sin^{2} \beta=1-(\frac{\dot {z}}{\kappa})^{2}$ holds for circular binaries, in which case Equation  \ref{accel2} defines an ellipse $ (\frac{P\ddot {z}}{2\pi\kappa})^{2}+(\frac{\dot{z}}{\kappa})^{2}=1$.
However, if $e>0$, the curve {provided by Equation \ref{accel2}}  will be {distorted}.
{Thus, fitting} radial velocity {and} acceleration data  with Equation \ref{accel2} {results in a new} test for the eccentricity of the \eCBSMBH.
{This can be useful} when broad-line centroids or peaks exhibit velocity shifts that {match} those expected by orbital motion but are {caused by varied} BLR illumination \citep{ 10.1088/0067-0049/187/2/416,10.1088/2041-8205/742/1/L12, 10.1051/0004-6361/201423555,  10.1088/0067-0049/217/2/26}. {Examining} the {relationship} between acceleration and radial velocity will {aid in the elimination of false binary candidates}.
Figure \ref{vradvsaccel} shows differences in the velocity--acceleration curves when comparing circular to elliptical motion of the secondary SMBH.
\citet{10.1088/0004-637x/789/2/140} {and} \citet{10.1088/0067-0049/201/2/23} measured the accelerations of binary SMBH candidates  by dividing velocity changes by the rest-frame time intervals between observations, which can be affected by orbital phase and period \citep{10.1088/0004-637x/789/2/140}. However, if the radial velocity curve can be folded over the photometric phase $\psi=\frac{t-t_{0}}{P}$, the following will be true: $dv_{\rm rad}/d\psi=PdV_{\rm rad}/dt$. {The last equivalence suggests} that without knowing the period $P$, the values of scaled radial acceleration {by period} ($PdV_{\rm rad}/dt$) can be obtained {simply} by computing the phase derivative of the curve.

\subsection{Refinement of binary  detectability  in contrast to other CBD emission phenomena}\label{sec:cbdrefin}
{Because} of a periodic variation in the mutual distance of the SMBHs, there could be a range of orbital phases where the sublimation ring is {totally contained in the  cavity of the CBD}. However, outside of this phase range, {the dust ring and CBD may intersect}. {If} the sublimation ring is {completely} inside the cavity, the emission may {emanate} from the intersection of the sublimation ring and the arms of the infalling matter from the CBD, {assuming} the CBD is sufficiently thin. The mass inflow from the CBD is greatest around the apoastron of the binary orbit, as demonstrated by \cite{10.1093/pasj/59.2.427}. The majority of infalling gas {is captured} by black holes shortly before the periastron, where we {would} expect the most of such emission to occur. 

{ Nonetheless}, when the CBD is sufficiently thick, we {anticipate} that optical and IR radiation will be{ released} mostly from the  outer regions of the CBD. {In reality}, the CBD imperfections may be more {complicated}, {resulting in more complex occurrences in inflows}.
{The Keplerian motion is not just a distinguishing feature of \eCBSMBH s\, in their early stages of evolution}.  {Such motions} can also be produced by disc spots \citep[if their velocities are Keplerian; see e.g.][]{10.1086/511032}. 
{Whether the angular velocity of a hot spot  is constant or not implies a circular or eccentric orbit in the disc \citep{10.1086/304460}.}
{It is also feasible} that the radial velocities of these non-static anomalies are not Keplerian \citep{10.1086/511032}.
{For \eCBSMBH\, detection}, it is {critical} to determine whether or not the photo-centre displacement is Keplerian. 

{Furthermore}, if the displacement {is caused by Keplerian motion}, the {inferred} orbital eccentricity of the  object {based on} the astrometric offsets could be used to {distinguish between} the \eCBSMBH\, and {another} phenomenon (bright spot, spiral) {that is} mimicking the signal.  {Bright} spots and spirals may have small eccentricities, {whereas the orbits of} \eCBSMBH s\,  {may be} more eccentric. { The fact that the hot spot can remain with nearly constant strength for multiple orbits before decaying over a shorter time \citep{10.1086/304460} allows us to rule out such emission as a binary possibility.
        The timescales related to one-armed spiral waves  and
        precession of the warp might be   too
        long to account for the observed orbital period  \citep{10.1086/304460}.}

{The above arguments are} not a prescription for {distinguishing between}  \eCBSMBH s\, {and other occurrences}, {but rather provide some ideas as to the various possible phenomena that may be seen in observations}.

\subsection{Refinement of binary detectability via Joint SpectroAstrometry and Reverberation Mapping}\label{sec:sarm}
{
        We highlight the possibility of combining SpectroAstrometry (SA) and Reverberation Mapping (RM; SARM) 
        for future detection of CB-SMBHs using GRAVITY+. In practice, GRAVITY measures the spatial distributions of ionised gas in the nuclear region via SA, whereas RM provides the most information on the radial distributions of the nuclear regions. As a result, SARM analysis will, in theory, reveal the global picture of nuclear regions first advocated for 3C 273 by \cite{10.1038/s41550-019-0979-5} and effectively applied to NGC 3783 by \citep{Shangguan2021}. Considering profile variations during the RM campaign, \cite{Li2021} developed more sophisticated method for two-dimensional SARM  which will improve measurements of SMBH masses and cosmic distances. The use of SARM with observations of CB-SMBHs in the future appears to be promising.}

\section{Discussions: models and measurements}\label{sec:issue}

Here we distill a list of potentially important issues for the interpretation of  CB-SMBH interferometric observations {arising because of model limitations (Section \ref{sec:modellim}) and challenges faced in measurements of binary radial velocity (Section \ref{sec:measurevel}).
Also, we infer the lower limit on {\eCBSMBH} mass based on radial velocity with a brief summary of radial velocity measurement methods that could be employed (Section  \ref{sec:masslim}).}

\subsection{\eCBSMBH\, models}\label{sec:modellim}
The fitted models to the AGN interferometric data need to be as straightforward as possible to avoid degeneracies \citep{10.1051/0004-6361/201527590}.
Despite the fact that the actual brightness distribution of a CB-SMBH can be quite complex,  the model given by \citet{10.3847/1538-4357/abc24f}\,  can provide a first-order approximation of the shape and size and serve as a building block for more complex geometries \citep[e.g. similar to the mid-infrared interferometry   of AGNs;][]{10.1038/nature02531,10.1088/0004-637X/806/1/127,10.1051/0004-6361/201527590}.
The technique used
by \citet{10.3847/1538-4357/abc24f} implies that periodicity associated with SMBHBs manifests in a Keplerian form.
However, there are indicators that a certain category of non-Keplerian periodic SMBHBs can exist \citep[see][]{10.1103/PhysRevD.101.043022}; for example, flaring, such as OJ 287 \citep{10.3847/1538-4357/aadd95}.
It should be noted that the \citet{10.3847/1538-4357/abc24f} method is designed to represent the brightness distribution as simply as possible without assuming any physical link (power law or otherwise) to the unresolved spatial scales.

The interaction between CBD material and the CB-SMBH provides instructive instances of the relationship between processes occurring at different scales.
Because of perturbations, matter in the CBD disc, for example, can traverse the gap in tiny streams, the eventual destinations of which depend on the precise angular momentum of the 
matter \citep{10.3847/1538-4357/aad8b4}. One scenario is that the binary torques thrust falling matter back, causing it to shock against the CBD; deflection in these shocks creates gas with substantially lower angular momentum, which plunges into the binary zone \citep{10.3847/1538-4357/aad8b4}. 
Accretion rates in the CB-SMBH system are another example of a phenomenon at a smaller scale that can influence the detection of these objects.
Periodic mass accretion rates can cause an overdense lump to form in the inner circumbinary accretion disc  \citep{10.1088/0004-637X/783/2/134}, which can mimic the astrometry signal.

Furthermore, because the spectral energy distribution of a circumbinary disc has a steeper power-law curve, accretion changes will be more noticeable at shorter wavelengths \citep{10.1038/nature14143}. {Another complexity of the binary--CBD interaction could be cycling transitions between type-1 and type-2 AGNs \citep[][]{10.1051/0004-6361/202039368}.} {In this scenario, both black holes are forming mini-discs around themselves by striping gas from the  inner edge of the circumbinary disc. The tidal torque caused by black holes on the mini-discs is strong enough to cause an exchange of angular momentum between the discs and the binary orbit. For retrograde mini-discs, tidal torque rapidly squeezes the tidal parts of the mini-discs into much smaller radii, causing higher accretion and short flares before the discs shift into type-2 AGNs.  Prograde mini-discs gain angular momentum from the binary and rotate outward, rapidly transitioning from type-1 to type-2 AGNs.}

Some specific occurrences in binary motion can cause the astrometric signal to be perturbed.
In the case of an eccentric binary, with different masses of components, the less massive black hole may get closer to the circumbinary disc than the larger one, tidally splitting gas from its inner edge \citep{10.1093/pasj/65.4.86} or exciting spiral density
waves. Such disturbances can cause the centre of mass of the circumbinary disc  to move and even produce an additional wobble in the secondary SMBH position, while time-varying, asymmetric light scattering by the disc can cause shifts in the photo-centre position.
Likewise, while beamed jet emission is expected to be associated with an individual black hole in a binary system, it is possible to encounter a non-thermal contribution from a precessing jet \citep{10.1117/12.460869}.

The consequences of finite sampling on eccentric ($e\gtrapprox 0.5$) RV curves can be anticipated. The RV curves  (see Equation \ref{radvel222}) seem flatter across a larger fraction of a period as the orbits become more eccentric (the binary component spends more time near apoastron). Because there is a greater chance of sampling RV data in these flat places (then at the peak), the observed RV curve may appear to be consistent with a constant velocity (no binary companion) even when numerous periods are sampled; unless the  peak in the RV curve is sampled as the binary component passes through periastron. 

Also, the RV data can be influenced by $\omega$ for higher eccentricities (Equation \ref{radvel222}).
For the circular binaries,  a small portion of the RV curve near maximum and minimum velocity has a flat slope and closely resembles a constant velocity (no binary), but a small portion near systemic velocity has a steep slope and would be easier to distinguish from a constant velocity, assuming a sufficient number of data points.

Astrometric data are similar to RV data in the sense that they are modified sine functions (e.g. see right panels in Figure \ref{model}). However, astrometric data are presented in two mutually orthogonal directions.
GRAVITY + data collected near the pericentre, where the gradient  varies quickly, will be better for eccentric binary model fitting than data collected near the apocentre. Because the binary component spends very little time near pericentre at high $e$ and small orbital period, sparsely sampled data may miss this key region of the orbit.

\subsection{Measurement of binary radial velocity  as a shift  of  a spectral line centroid wavelength}\label{sec:measurevel}

While double-peaked broad lines are {unlikely} to be a useful diagnostic of SMBHBs \citep{2000SerAJ.162....1P,10.1088/0004-637X/725/1/249, 10.1016/j.newar.2011.11.001, 10.1007/s10509-015-2647-2,10.1093/mnras/stab1510}, single-peaked broad-line offsets can be analysed
\citep{10.1088/0067-0049/201/2/23}. The probability of one component being active is {substantially} higher than the probability of both components being active {at the same time}, and t{he permitted} binary parameter space is likewise larger than in the case of double-peaked broad lines \citep{10.1088/0004-637x/789/2/140}. 
Monitoring campaigns are {unlikely to be able} to record several cycles of radial velocity curves from \eCBSMBH s. {As a result,} the signature of a binary will be monotonic (increase or decrease) or even flat in the observed radial velocity \citep[see ][and their Figures 3, 4, and 5]{10.1093/mnras/stx452}, {whilst} the spectral lines will oscillate around their rest centroid wavelengths by  $V_{\rm rad}/c$.

{The} spectral line will be single-peaked if the secondary SMBH has dominant BLR radiation. Radial velocity can be expressed as a wavelength shift ($\Delta \lambda$) in a spectral line centroid wavelength  $\lambda$ as {follows}:

\begin{equation}
\frac{\Delta\lambda}{\lambda}=\sqrt{\frac{R_{\rm g}}{\tilde{a}}}\frac{1}{1+q}\frac{\sin i}{\sqrt{1-e^{2}}}\left[\cos(f+\omega)+e\cos{\omega}\right], 
\label{ff}
\end{equation}
\noindent where $R_{\rm g}=G(M_{1}+M_{2})/{c^{2}}$ {denotes} the gravitational radius of the  binary and $q=M_2/{M_1}$.
Even if the line profile is perturbed, the periodic {wobbling} will be imprinted and may still be observable, as shown above.

{For} observer inclination $i=90^{\circ}$,   $\Delta\lambda/\lambda$  reaches maximum value. {Under such geometric constraints}, the amplitude at $1000 R_{\rm g}$ will be {approximately}  $10^{-0.5}$  {less} than the amplitude at $100R_{\rm g}$.
If {the term} $\cos(f+\omega)\sim 0$ {is valid}, the amplitude will be multiplied by $e\cos\omega$; {whereas}, {when} $\cos(f+\omega)\sim 1$ {holds,} the amplitude will be multiplied by a factor of $(1+e\cos \omega)$. 

\subsection{Measurement of the lower limit on  binary mass from radial velocity}\label{sec:masslim}

Until now, the velocity curves have been produced by a few long-term spectroscopic monitorings \citep{10.1088/0067-0049/201/2/23,10.1088/0067-0049/221/1/7,10.1093/mnras/stx452}.
\citet{10.1093/mnras/stx452} {estimated} the radial velocities of \eCBSMBH\, candidates with broad (single-peaked)  H${\alpha}$ or H${\beta}$ lines and 
offsets of  $|\Delta V|> 1000 \mathrm{km\,s}^{-1}$ \citep{10.1088/0067-0049/201/2/23}. 
{It has also been shown} that long-term radial velocity curves can be fitted to get constraints on orbital elements \citep{10.1093/mnras/stx452}.
{We expect } that the amplitude increases with { binary total mass $M_{1}+M_{2}$} (see e.g. Equation \ref{radvel222}).
{Taking these constraints into consideration we can infer  the lower limit on binary mass as 
}

\begin{equation}
M_{1}+M_{2}>\frac{(1000 \mathrm{km\, s}^{-1})^{2} \tilde{a}(1-e^{2})}{G \sin^{2}i\left[\cos(f+\omega)+e\cos{\omega}\right]^{2}},
\end{equation}
or {if the secondary is located} in pericentre  $f=0,$
\begin{equation}
M_{1}+M_{2}>\frac{(1000 \mathrm{km\, s}^{-1})^{2} \tilde{a}(1-e)}{G \sin^{2}i\cos\omega (1+e)}.
\end{equation}
{ Here, we provide a quick summary of the methodologies for radial velocity measurement that could be employed and eventually upgraded in the context of the concerns discussed above.}
{Observational} searches for close binary SMBHs using single broad-line spectroscopic spectra can be {divided} into several categories. The first type has targeted the quasars with broad lines located at their systemic velocities 
\citep[that would be binaries in conjunction; ][]{10.1088/0004-637X/777/1/44, 10.1088/0004-637X/775/1/49,10.3847/1538-4357/834/2/129}. {The} second type {of survey} targets sources {with}  broad emission lines {that} are offset from the rest frame by thousands of $\mathrm{km}\,\mathrm{s}^{-1}$  \citep{10.1088/0004-637X/738/1/20, 10.1088/0067-0049/201/2/23,   10.1093/mnras/stt831, 10.1088/0004-637x/789/2/140}. 
{In most cases}, two spectra taken years apart {were used} to measure or constrain the  radial velocity fluctuations of binary candidates. These methods take into consideration {the fact that}, for example, the  $H{\alpha}$ narrow line (NL) is assumed to be at zero velocity. The velocity of a {displaced}  peak of the $H{\alpha}$ broad line (BL) is {given by \citet{10.3847/0004-637x/817/1/42} } 
\begin{equation}
V=\left(\frac{\lambda_\mathrm{BL}-\lambda_\mathrm{NL}}{\lambda _{\mathrm{NL}}}\right)c.
\end{equation}

\section{Outlook for the future}\label{sec:prospect}

{
        We describe the potential of using interferometry to measure  the angular position of the photo-centre at the emission line of {\eCBSMBH}s which could be useful for GRAVITY+ successors, and {\eCBSMBH} relevance for nano-Hz gravitational wave astronomy.}

\subsection{Centroid measurements}

When {employing} spectro-astrometry to {determine} the  origin of a certain emission line, the source position should be {precisely} mapped by taking into {consideration} the centroid of the continuum emission. The approach provided here for determining the angular position of the photo-centre at the emission line might be useful for GRAVITY+ successors. {The} intensity ratio of the continuum and emission line  is used to weight the extent of the emission line region offset.    Estimates of associated Keplerian velocities can be {used to make a} preliminary {determination} of whether the emission line originates near the secondary or from a CBD. {The corresponding} Keplerian velocities for CBD and the active secondary,{ respectively}, are $\sqrt{\frac{G(M_{1}+M_{2})}{2\tilde{a}(1+e)}}$ {and}
$\sqrt{\frac{G(M_{1}+M_{2})}{a_2}}$, where $a_2$ is the barycentric distance of the  secondary. {Their straightforward comparison shows} that the Keplerian velocity at CBD distance would {have been around}
$\sim (2(1+q)(1+e))^{-0.5}$ times that of the active secondary component. However, if the full width at half maximum  (FWHM) of observed emission line spectra is {substantially} larger than the expected Keplerian velocity associated with CBD, this {indicates} that the emission line is {emerging} from the disc {surrounding} the secondary. 

If an emission line {comes} from the BLR region bound to the secondary, then our raw centroid measurements will {include} both  emission line and the continuum centroids:

\begin{equation}
\mathbf{C}=\frac{F^{\rm l}\mathbf{r^{\rm BLR}}}{F^{\rm l}+F^{\rm c}}+\frac{F^{\rm c}\mathbf{r^{\rm c}}}{F^{\rm l}+F^{\rm c}}
\label{eq:centroids}
,\end{equation}
\noindent
\noindent where $F^{\rm c}$ and $F^{\rm l}$ are {the} continuum and line fluxes, {respectively, and}  $ \mathbf{r^{\rm c}}$ $\mathbf{r^{\rm BLR}}$ are {their corresponding} locations.
{Notably,} the centroid of the emission line (the first term) naturally vanishes for $F^{\rm l}\rightarrow 0$.
Similarly, the  the distance between the  centroid positions and secondary $\mathbf{r}{_2}$\, {is as follows}:
\begin{equation}
\mathbf{C}-\mathbf{r}{_2}=\frac{(\mathbf{r^{\rm BLR}}-\mathbf{r}{_2})}{1+F^{\rm c}/F^{\rm l}}+
\frac{(\mathbf{r^{\rm c}}-\mathbf{r}{_2})}{1+F^{\rm l}/F^{\rm c}}
\label{eq:shortcent}
.\end{equation}
{When the emission line flux is weak} ($F^{\rm l}\rightarrow 0$), the centroid separation {coincides with} the continuum separation, that is,  $\mathbf{C}-\mathbf{r}{_2}\sim{\mathbf{r^{\rm c}}-\mathbf{r}{_2}}$.
Otherwise, when {the contribution of the emission line is substantial}   ($F^{\rm l}>>0$) and  the time lag  $\mathbf{r^{\rm BLR}}/\mathbf{r^{\rm c}}\propto\tau$ {is assumed}, the individual term $\mathbf{r}{^c}/ \mathbf{r}{^c}$ can multiply  the right  side of Equation \ref{eq:centroids}, {yielding the}   relation
\begin{equation}
\mathbf{C}\propto \mathbf{r^{\rm c}}\frac{\tau F^{\rm l}+F^{\rm c}}{F^{\rm l}+F^{\rm c}}.
\end{equation}
{ Because the discs can be brighter on one side than the other, temperature variations can arise along} the inner and outer edges of the CBD and the SMBH disc  \citep{10.1088/0004-637X/785/2/115}. {However,} the CBD can be  hotter than the {binary component disc  by a factor of two}, but not as hot as the  innermost regions of the disc of the binary component   \citep{10.1093/mnrasl/slu075}.
{However}, if we relax {the assumption} that there is a sharp surface density cut-off at these boundaries, and assume that both the dust ring and CBD emit radiation at distances of $\mathbf{r}^{\rm sub}$ and  $\mathbf{r}^{\rm CBD}$, respectively, the expression for the photo-centre offset with regard to the secondary is as follows:

\begin{equation}
\mathbf{C}-\mathbf{r}{_2}=\frac{(\mathbf{r^{\rm sub}}-\mathbf{r}_{2})}{1+F^{\rm CBD}/F^{\rm sub}}+
\frac{(\mathbf{r^{\rm CBD}}-\mathbf{r}_{2})}{1+F^{\rm sub}/F^{\rm CBD}}
\label{eq:centrsubcbd}
.\end{equation}

{The second term on the right} side of Equation \ref{eq:centrsubcbd} {could be accounted for as in the case of} the dust ring and CBD intersection, {so that} the centroid offset with regard to the secondary (i.e. $\mathbf{C}-\mathbf{r}_2$) {is augmented by the CBD contribution to total emission
:} 
\begin{equation}
\frac{F^{\rm CBD}}{F^{\rm CBD}+F^{\rm sub}}.
\end{equation}
{If  the inequality}  $\mathbf{r}_{\rm CBD}>\mathbf{r}^{\rm sub}$ {is true},  the emission will be {dominated by} CBD, and the first term on the right  side of Equation \ref{eq:centrsubcbd}  will vanish, {implying that} $\mathbf{C}\sim \mathbf{r}^{\rm CBD}$ {holds}.  In reality, the continuum emission could come through the emission line channel. {In this case}, the emission line {centroid offset} can be {calculated}  by subtracting the continuum astrometric signal ({the} second term {on} the right side of  Equation \ref{eq:centrsubcbd}) from the observed signal (the term on the left side). {This indicates} that the photo-centre {of a} line will be displaced with respect to the photo-centre of the continuum. This shift reflects the fact that the photo-centre of continuum emission is shifted towards the CBD {rather than} the location of the secondary SMBH{}.

If the {emission line comes} from the disc-like BLR of the primary SMBH, but the continuum {is emitted} from the dust ring surrounding the secondary, we can {express} the raw measurement of the astrometric centroid as:

\begin{equation}
\mathbf{C}=\frac{F_{1}^{\rm l}(\mathbf{r_{1}^{\rm BLR}+\mathbf{r_{1}}})}{F_{1}^{\rm l}+F_{2}^{\rm c}}+
\frac{F_{2}^{\rm c}(\mathbf{r_{2}^{\rm c}}+\mathbf{r_2})}{F_{1}^{\rm l}+F_{2}^{\rm c}},
\label{eq:centroids2}
\end{equation}
\noindent where $\mathbf{r_1}$ and  $\mathbf{r_2}$ are {the} positions of the primary and secondary SMBH, $\mathbf{r_{1}}^{\rm BLR}$, {respectively},  and 
$\mathbf{r_{2}}^{\rm c}$ are {the} positions of line and continuum emissions  with respect to {the} primary and secondary.
{Another possibility is that} the continuum emission {can} be extracted ({by} subtracting   the second term on the right side of Equation \ref{eq:centroids2} from the total centroid),  {in which case we may also} measure the centroid of the line  (the first term on the left-hand side).
The photo-centre of the emission line will be displaced from the axis {defined by the} primary and secondary SMBH {positions}; {however} the centroid of the continuum will {be} anchored to this axis. {The photo-centre of an emission line} will clearly be closer to the primary SMBH, which {is supposed to be} a line-emitting source.{ It is worth noting} that Equation  \ref{eq:centroids2} is {written} in a barycentric coordinate system, {implying} that the calculation procedure must include the barycentre as {an} unknown parameter. 

\begin{figure}[ht!]
        \includegraphics[ width=0.5\textwidth]{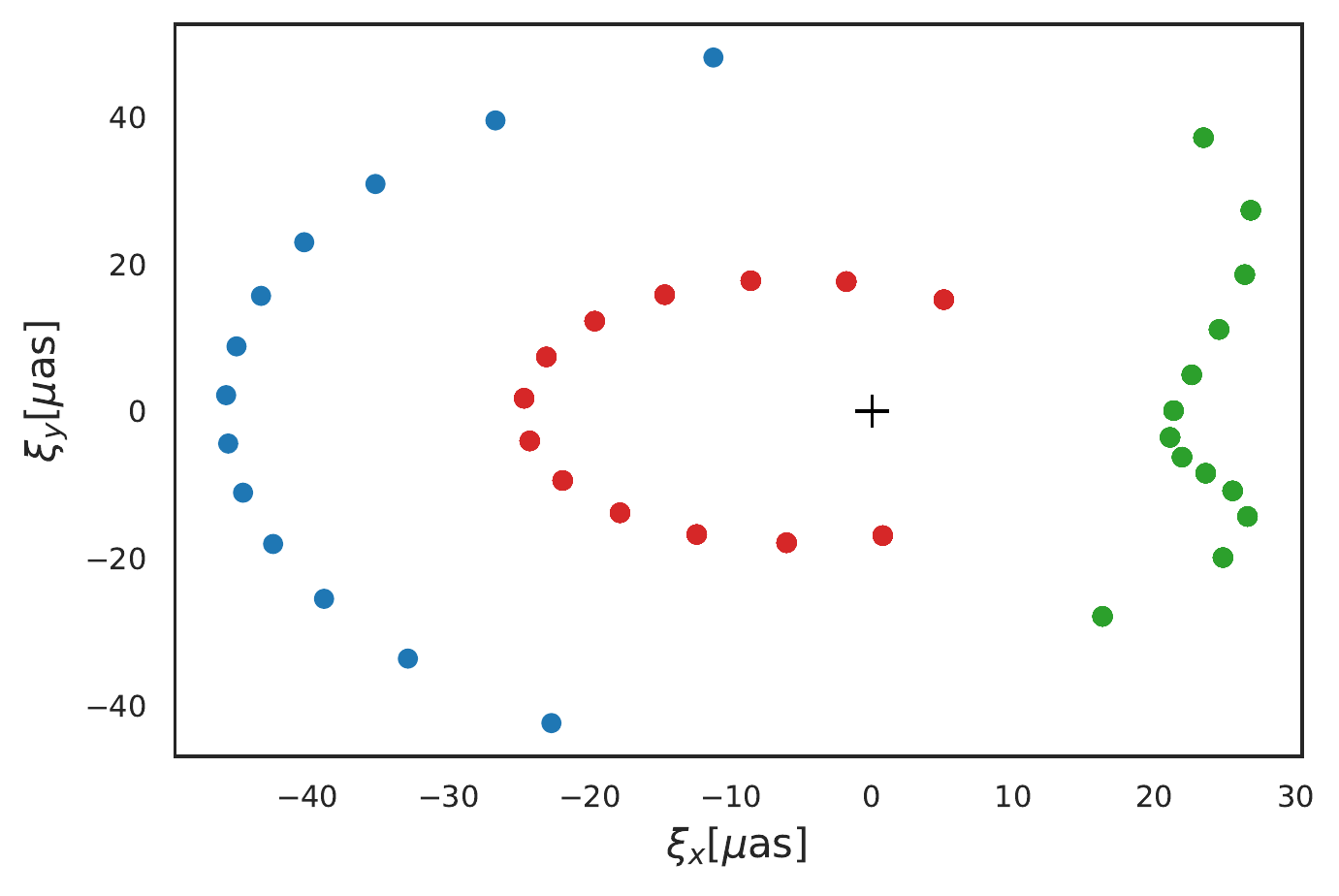}
        \caption{ {Simulation of } photo-centres for  continuum source (blue dots), emission line (red dots), and relative positions  (green dots) of emission line photo-centre with respect to the continuum photo-centre. {The binary system  parameters are:  $M_{1}=6\times 10^{7} M_{\odot}, M_{2}=4\times10^{7} M_{\odot},e=0.5,\Omega_{1}=180^{\circ}, \Omega_{2}=100^{\circ},\omega_{1}=0.1^{\circ},\omega_{2}=180.1^{\circ}$,  $\Rsub=150 \mathrm{ld}$. The mean mutual distance of the components is 100 ld, the binary  distance is $\sim 700$ Mpc, and the observer position angle is $40^{\circ}$}.
                The emission line source is located in the BLR {that is} bound to the primary, {whereas} the continuum is defined as the barycentre of the dust ring arc that is bound to the secondary and split by the CBD. The cross {represents} the barycentre of the \eCBSMBH.}
        \label{emline}
\end{figure}

However, {if} Equation  \ref{eq:centroids2} {is rewritten} relative to the primary component {as follows}:
\begin{equation}
\mathbf{C}-\mathbf{r_1}=\frac{F_{1}^{\rm lr}\mathbf{r_{1}^{\rm BLR}}}{F_{1}^{\rm l}+F_{2}^{\rm c}} +\frac{F_{2}^{\rm c}\mathbf{r_{2}^{\rm c}}}{F_{1}^{\rm l}+F_{2}^{\rm c}}+ \frac{F_{2}^{\rm c}(\mathbf{r_{2}}-\mathbf{r_{1}})}{F_{1}^{\rm l}+F_{2}^{\rm c}}
\label{eq:centroids22}
,\end{equation}
\noindent {then}  extracting the centroid of the line (the first term on the right  side) {becomes more difficult because,}  {in addition to} the photo-centre of the continuum (the second term), there is a third term related to the relative positions of the primary and secondary {components}. For example, if an emission line  originates at 70 ld from the primary, the angular separation between the source and primary at the distance of 700Mpc {is} $\sim 20\mu$as, which can potentially be measured. 
{ To make it easier to assess these effects, we generated} 13 artificial  observations of the position of the three  photo-centres (see Figure \ref{emline}). 
The red dots in represent the photo-centre of the emission line originating from the primary BLR, whereas the blue dots represent continuum photo-centres computed as centroids of dust rings bound to the secondary. We can observe that any triplet of the CB-SMBH barycentre (cross), emission line, and continuum photo-centre  is non-collinear. As a result of non-collinearity, the relative position of the emission line centroid with respect to the continuum centroid will have an asymmetric `wavey' shape (green dots).

Finally,  photo-centre displacements { determined by} spectro-astrometry ($\mathbf{C}$) are mathematically equivalent  to   phase ($\mathbf{\phi}$) {determined by} spectro-interferometry, and {therefore the following relationship holds} \citep{10.1088/0004-637X/752/1/11}:

\begin{equation}
\mathbf{ \phi}=-\frac{2\pi \mathbf{C}}{\sigma}
,\end{equation}
\noindent where  $\sigma $ is  the FWHM.
{ To demonstrate the above formula, consider} an unresolved source $\frac{(b1,b2)}{\lambda}\cdot (\alpha,\delta)  < 1$, where $\frac{(b1,b2)}{\lambda}$ is the interferometric baseline vector,   $(\alpha,\delta)$ is the source  position on the sky,  and the phase is proportional to the photo-centre of the source projected onto the baseline   \citep[see][and references therein]{10.1051/0004-6361/202038733} $\phi\sim -2\pi (\frac{(b1,b2)}{\lambda})\cdot\mathbf{C}$.
If {the unresolved source is an} AGN {with} two components of emission, the continuum emission from hot dust and the line emission from the BLR, the centroid of brightness of the system will be
\begin{equation}
\mathbf{C}=\frac{F^{\rm l}}{F^{\rm l}+F^{\rm c}}\mathbf{x}^{\rm l}+  
\frac{F^{\rm c}}{F^{\rm l}+F^{\rm c}}\mathbf{x}^{\rm c}.
\end{equation}
To calculate the differential phase, we subtract the continuum emission location ($\mathbf{x}^{c}$)  from both sides of the preceding equation and multiply both sides by a factor of $-2\pi\frac{(b1,b2)}{\lambda}$ as follows:

\begin{equation}
\Delta \phi=\frac{F^{\rm l}}{F^{\rm l}+F^{\rm c}}(\phi^{\rm l}-\phi^{\rm c}).
\end{equation}

Similar {information} can be obtained {by using} the Fourier transform to relate visibility and object brightness distribution, as well as  Fourier transform properties. The Fourier phase stores deviation from centre-symmetry, { which is one of these features} \citep{2008LNP...742..123W}.  
{For example,}  an angular shift of $(\alpha,\delta)$ of the brightness distribution {results in} a visibility phase shift of $\phi = {(b1,b2)\cdot(\alpha,\delta)}/{\lambda}$ \citep{2008LNP...742..123W}.

\subsection{{\eCBSMBH} in the context of gravitational waves}

        The anticipated eccentricities and eccentricity evolution for SMBHBs is discussed above (Section \ref{sec:intro}). How this relates to the signals that could be detected by  pulsar timing arrays (PTAs) is briefly mentioned here.
        The expected timescales  of AGN electromagnetic activity are
        much longer than those of their gravitational-wave channel.
        Mergers of black holes of similar masses could be one route for generating SMBHs found in quasars \citep{10.1103/PhysRevD.104.024027}.
        In such case,  the number of
        mergers between black holes with masses larger than
        $10^6 M\odot$ would be between 1 and 10 \citep{10.1088/0264-9381/26/9/094033}. 
        {Moreover,  the number of galaxy mergers harbouring SMBHs could be a few, as data imply that this layout may be sufficient to explain the presence of SMBHs in quasars \citep{10.1126/science.1184246, 10.1103/PhysRevD.104.024027}.}

{Gravitational radiation} enters the low {nano-Hz}
frequency band at sub-parsec length scales of CB-SMBH mergers, where it is strong enough {to be detected by}  PTAs  \citep{10.1103/PhysRevD.92.063010}.
The International PTA (IPTA) consortium combines resources  from multiple PTAs in order to identify nano-Hz GWs more quickly  \citep{10.1088/0264-9381/27/8/084013,  10.1093/mnras/stz2857}.

{In theory}, continuous GW detection {using} PTAs {might offer} orbital frequency and eccentricity measurements {for the SMBH binary} system  \citep{10.1007/s00159-019-0115-7}.
However, chirp mass and source distance cannot be {determined} directly {until} the orbital frequency evolution is detected during the PTA observations or the host galaxy of the continuous-wave source is identified \citep{10.1007/s00159-019-0115-7}.
{The} North American Nanohertz Observatory for Gravitational Waves (NANOGrav) {discovered} an {unusual} low-frequency signal in data collected over a 13-year {period}   \citep{10.3847/2041-8213/abd401}. 
{Under the assumption that} the signal is {genuinely} astrophysical, the potential implications for the binary population of SMBHs were investigated by \citet{10.3847/2041-8213/abd401}.

{
        Even without invoking a third perturbing SMBH,  stellar `loss cone' scattering can increase binary eccentricity. This has been seen in some 
        numerical simulations \citep{10.1016/S1384-1076(96)00003-6,10.1088/0004-637X/719/1/851},  where it appears that equal-mass binaries (essential for PTAs) with extremely low initial eccentricity would either retain {eccentricity} 
or grow somewhat more eccentric.  At any starting eccentricity, for binaries with extreme mass ratios, the eccentricity can evolve considerably, allowing high values to be maintained even into the PTA range
        \citep{10.1088/0004-637X/719/1/851,10.1088/1742-6596/363/1/012035}.}
\citet{10.3847/2041-8213/abb2ab} found that the  timescales to coalescence of  spectroscopically selected, subparsec, circular 
binary  candidates \citep{10.1088/0067-0049/201/2/23} are in the range of
$10^{4}-10^{9}$yr, assuming that their orbital evolution in the PTA frequency band is driven by emission of gravitational waves. 
For PTA, the eccentricity $e=0.5$ of CB-SMBHs is {a threshold} where the difference 
between eccentric and circular signal models is greatest \citep{10.3847/0004-637X/817/1/70}.

{The orbital eccentricities of SMBHBs generating nano-Hz GWs can be rather large \citep{10.1007/s00159-019-0115-7, 10.1093/mnras/stx1638}. Due to GW emission, such CB-SMBHs will coalesce in less than a Hubble time, depending on the orbital eccentricity \citep{PM63,P64}. }
In very-steep-profile galaxy mergers, SMBHBs with eccentricities of  $0.4-0.6$ and very short coalescence times of $\sim0.4 \mathrm{Gyr}$ are found  \citep{10.1088/0004-637X/732/2/89, 10.1088/0004-637X/749/2/147}.
The dependence on eccentricity of the coalescence time under gravitational wave emission $T_{\rm {coal, GW}}$ in such mergers may scale as   $T_{\rm {coal, GW}}\sim (1-e^{2})^{3.5}$ \citep{10.1088/0004-637X/749/2/147}.

{Theoretically, it is expected that eccentricity and the Keplerian orbital frequency of binaries co-evolve in a 
        mass-independent way \citep{10.3847/0004-637X/817/1/70}.
        For example, if a binary starts with an eccentricity of $e_{0}=0.95$ at an initial epoch and its orbital frequency is 1 nHz, it will reach  $e\sim 0.3$  by the time its orbital frequency reaches 100 nHz \citep{10.1007/s00159-019-0115-7}.
        SMBHBs in eccentric orbits, as in the blazar OJ 287 \citep{10.3847/1538-4357/aadd95}, are promising nanoHz GW sources for the rapidly maturing PTA  efforts \citep[PTA][]{10.1088/0264-9381/30/22/224008,baas.aas.org/pub/2020n7i195}, and 
         \citet{10.1103/PhysRevD.101.043022} devised an accurate and effective prescription for obtaining PTA signals caused by
        isolated SMBHBs inspiraling along general relativistic eccentric orbits.}

\section{Conclusion}
{
In this work, we predict the performance of GRAVITY + observational campaigns  regarding the detection and analysis of eccentric binary SMBHs ({\eCBSMBH s}). 
It is commonly assumed that, to extract  the signal from such objects, an observational timeline much longer than the orbital period of the  object is needed.
Because of the lack of data  for {\eCBSMBH s}, we simulate two example cases of GRAVITY +  campaigns, each  with a different total number of observations (N), a different time baseline  ($\mathcal{T}$), and different   \eCBSMBH\, dynamical parameters.
We show that when both GRAVITY+  accurate astrometric and high-precision RV  measurements are available, it is possible to detect the basic orbital parameters ($M, P, e$) of 
{\eCBSMBH s} with observational timelines that are considerably shorter than their orbital period.
Based on current GRAVITY + instrument information, we constructed mock astrometric and RV  observations of {\eCBSMBH s} using the binary dynamical model (with constant, evolving, and fluctuating hot dust emission) as a representative  of our current understanding of these eccentric SMBHBs.
We then examined the detectability of {\eCBSMBH s}\,   using the 
 photo-centre offset caused by the intersection of the dust ring of the secondary SMBH  and the CBD;  the astrometric signal in the limit of binary  eccentricity as a main factor of {\eCBSMBH} orbital shape; and Bayesian inference of {\eCBSMBH}\, basic orbital parameters ($M, P, e$) from GRAVITY + mock campaigns covering $5\%$ and $10\%$ of {a} whole orbital period of {the source}.
}

Based on these considerations, we reach the following  conclusions:
\begin{itemize}
\item{{We estimate that the expected number of   {\eCBSMBH s}  within reach of  GRAVITY+  inside a sphere of $z<0.3$ is between approximately 4 and 13, assuming an arbitrary range of eccentric binary masses $M$, eccentricities ($e$), and periods ($P$).
In addition to the above blind estimate, using the  expected distributions of SMBHBs per $\log z$, the quasar luminosity function, and assuming that $\sim 10^{-3}$ is the fraction of CB-SMBHs in local bright AGNs, we  obtain a similar prediction that GRAVITY+ could discover about $7$ {\eCBSMBH s}. }

}
        \item{
                {Using} the GRAVITY+ wavelength detection limit, we {assess} the detectability of evolving hot dust emission in a \eCBSMBH\, {system}. 
                { We compute} the minimum value of flux ratios of NIR emission originating in the dust ring to the optical continuum as a function of dust ring dimension for different photo-centre displacements.
                At an astrometric observing wavelength of  $\sim 2 \mu$m, we predict that GRAVITY and GRAVITY+ could detect non-static hot dust emission of {an {\eCBSMBH}}. However, there {may be a} band of parameter space where some dust emission {goes undetected}.
        }
        \item{
                We  find analytic expressions for the S/N  for astrometric observations in
                the limit of  single-epoch astrometric  error, and {\eCBSMBH\, {parameters}:}  mass ratios,  periodicities (semi-major axes), eccentricity, {the arc of the observed binary orbit, and observation loss}.
                {
                        The width of pericentre passage is $\sim (1-e)^{2}P$, implying that 
 the enhanced velocity amplitude and acceleration near the periastron boost detectability in long-period objects, whereas   eccentricity makes detection more challenging at short periods, because uneven sampling frequently results in poor phase coverage during  rapid pericentre passage. 
                {When the pericentre width observational coverage is $\sim 10\%$},   we {predict}  that {\eCBSMBH s} with a binary mass ratio of $q=0.1$ at a mutual distance of 0.01 pc would be impossible to detect. However, we demonstrate that the S/N of {\eCBSMBH s} and their detectability increase with increasing $q$ and with better observational coverage of the pericentre passage.
                Also, we  mapped the  expected detection distance of {\eCBSMBH s} for different  mass ratios and semi-major axes across V magnitudes.}
                }

        \item{
                {We adopt the} Bayesian inference approach to extract {\eCBSMBH} signals from {GRAVITY+ simulated} campaigns, because resolving the {\eCBSMBH} orbits is a  non-linear problem. 
                {The results of the tests of orbital solutions} indicate that, with observational timelines covering at least  $\sim 0.1 P$,  models with evolving  dust   are more likely to identify the period, eccentricity, and the total mass of an \eCBSMBH.
        {When the observational coverage is low ($\sim 0.05P$), the quality of the detected period decreases, resulting in a relative discrepancy between the detected period and the true value of around 30 percent, whereas other parameters are unaffected.}
            {In general, Bayesian inference can improve the time efficiency of the GRAVITY+ observations}.
            }
        \item{{Specifically, we describe a method for measuring the angular position of the photo-centre at the emission line of an {\eCBSMBH}, which we expect will be useful for GRAVITY + successors.}}
\end{itemize}

{Additionally, we analysed  the effects   of various eccentric  binary mass ratios on signal and show that for  incomplete binary orbits  (observational coverage of $5\%-10\%$), Bayesian inference,  which joins multiple observational techniques,  generates more information on the system, either in a narrower posterior parameter density,  or in the potential if we include additional parameters in the model 
We addressed how to independently test the eccentricity of the \eCBSMBH\, based on radial velocity and acceleration data, and also  provide  a few possibilities that may be met in GRAVITY + observations that could imitate detection of an {\eCBSMBH}. Finally, we propose that the SARM technique \citep{10.1038/s41550-019-0979-5,Shangguan2021, Li2021} can assist GRAVITY + in binary detection, either as a follow-up or as an independent binary detection tool.}

\begin{acknowledgements}
AK acknowledges the funding provided by the Faculty of Mathematics University of Belgrade (the Ministry of Education, science and technological development of Republic Serbia contract 451-03-68$/$2022-14$/$200104).
AK and L{\v C}P thank the support by  Chinese Academy of Sciences President's International Fellowship Initiative (PIFI) for visiting scientist.
JMW and YYS thank the support from the National Key R\&D Program of China 2016YFA0400701, NSFC(-11991050,\,-11991054,\,-11833008). L{\v C}P acknowledges the funding provided by  Astronomical Observatory (the Ministry of Education, science and technological development of Republic Serbia contract 451-03-68$/$2022-14$/$ 200002).

\end{acknowledgements}

%
%

\bibliographystyle{aa}
\bibliography{akovacevic}

\begin{appendix} 
\section{Astrometric offset for the evolving hot-dust model in the context of the \eCBSMBH\, system }
\citet{10.3847/1538-4357/abc24f} demonstrated that the astrometric offset for the geometric evolving continuum model can be connected to the centroid of the intersection of two circles. Using the generic geometrical considerations discussed below, we demonstrate how this notion can be connected to the configuration of the \eCBSMBH.

{Let parametric} curve $\mathcal{C}$ describe an arc of a ring {with} constant density g(x,y,z) and cross-sectional area A(x(t),y(t),z(t)) {as shown below:}
\begin{eqnarray}
x = x(t),\\
y = y(t),\\
z=z(t).
\end{eqnarray}
Because density and area are constant along arc L, the centroid and centre of mass coincide, resulting in
\begin{eqnarray}
\mathbf{C}=  \frac{ \int \mathbf{r} dm}{m},
\end{eqnarray}
where  $dm=g(\mathbf{r})A(\mathbf{r})dL$ denotes the mass of an infinitesimal element in a ring.
We can rewrite the centroid equation as follows, assuming $A= 1$ for {ease of calculation:
        \begin{eqnarray}
        \mathbf{C}= \frac{\int \mathbf{r} g(\mathbf{r})A(\mathbf{r})dL}{\int g(\mathbf{r})A(\mathbf{r})dL}\sim \frac{\int\mathbf{r} g(\mathbf{r})dL}{\int g(\mathbf{r})dL.}
        \end{eqnarray}
        Finally, the centroid of an arc of a curve $\mathbf{r}(t)$ within the finite parameter interval} $t \in[a,b]$ can be represented as follows:
\begin{eqnarray}
\mathbf{C} = \frac{1}
{\int^{t=b}_{t=a}g(\mathbf{r})\cdot{\|\partial_{t}\mathbf{r}(t)}\|dt}\cdot\int^{t=b}_{t=a} \mathbf{r}\cdot g(\mathbf{r})\cdot\|\partial_{t}\mathbf{r}(t)\|dt.
\end{eqnarray}

{{Depending on the curve parametrisation}, this general equation will take on multiple forms. For example,} defining a circular arc in polar coordinates {yields} the arc centroid in its most compact form.
{ Here, we assume that}  the x-axis is {the} axis of the symmetry of the arc, and that the arc has radius $r$, and a central angle $2\alpha$.
Taking {into account} that the differential element of arc length is 
$dL=rd\theta$, the length of the arc is $L=2\alpha r$, and integration limits are $(-\alpha,\alpha)$, {we may calculate the centroid } $C_{X}=\frac{r \sin\alpha}{\alpha}$ \citep[see also][]{10.3847/1538-4357/abc24f}.
{We now present the astrometric offset of the evolving hot-dust emission model in the general concept of the \eCBSMBH\, system.}
Let $M_1$ and $M_{2}$ {be the} SMBHs loci,  points $P_1$ and $P_2$ {be the} intersections of the CBD and dust ring  bound to the secondary, {and} $P_3$ {be} the intersection of the line $M_{1}M_{2}$ and dust ring outside of the CBD. The barycentre of that arc of the dust ring is {provided by:}

\begin{equation}
\mathbf{C}=\overrightarrow{M_{2}P_{3}}\cdot \mathrm{sinc}{\angle (\overrightarrow{M_{2}P_{3}}, \overrightarrow{M_{2}P_{2}}})
\end{equation}
where  
\begin{equation}
{\angle (\overrightarrow{M_{2}P_{3}}, \overrightarrow{M_{2}P_{2}}})=
\mathrm{atan2}\left(\frac{\mathbf{L}_{\rm sub}}{\left\Vert \mathbf{L}_{\rm sub}\right\Vert}\cdot(\overrightarrow{M_{2}P_{3}} \times\overrightarrow{M_{2}P_{2}}),\overrightarrow{M_{2}P_{3}}, \cdot\overrightarrow{M_{2}P_{2}}\right),
\end{equation}

and $\mathbf{L}_{\rm sub}$ is the orbital angular momentum of the dust sublimation surface. \footnote{The $atan2$  is variant of $atan$ function that takes two arguments to be able to determine  the output angle in correct quadrant.}
{Because} both the dust ring and the CBD are circular, the dust ring  arc barycentre $\mathbf{C}$ is {placed on the line}   $M_{1}M_{2}$ as a bisector of angle $\angle (\overrightarrow{M_{2}P_{1}}, \overrightarrow{M_{2}P_{2}})$, {at any point of the \eCBSMBH\,  orbit}. {The developed} formula is {applicable to} both eccentric and circular CB-SMBH configurations. 
{Because of the features} of the sinc function, there is no difference in centroid position {whether} the motion of the dust is clockwise or anticlockwise, implying that it is independent of $\mathbf{L}_{\rm sub}$ orientation.

\section{Approximation of fluctuating hot dust continuum offset}

{We now establish formulations for astronomical offset, as indicated in the third branch of  Equation \ref{rsub}, for fluctuating the dust continuum model.}
Suppose {that only} continuum emission from the sublimation radius {is taken into account, ignoring}   CBD. In that case, the continuum flux offset (relative to the secondary or even barycentre of the CB-SMBH) will correspond to the dust ring position. 
{There are some empirical inferences regarding the dimension of the hot dust ring.} 
For example,  \citet{10.1088/0004-637X/788/2/159} discovered that the dust reverberation radius of a sample of 17 Seyfert galaxies is four to five times {greater than} their BLR radius and typically {a factor of two lower than the equivalent} interferometric radius. {Additionally}, the BLR radius determined by reverberation mapping is {less} than that determined by NIR interferometry \citep[see][and references therein]{10.1093/mnras/sty200}. However, {some exceptions are found in the literature. For example,} the dust radii of  NGC 4151 \citep{10.1086/507417, 2008A&A...486...99S},  Mrk 335 \citep{2014ApJ...782...45D},  {and} NGC 4593 \citep{2013ApJ...769..128B} {are approximately ten times larger} than the respective BLR radii  \citep{10.1088/0004-637X/788/2/159}. {These considerable} differences in dust radius and BLR radius indicate fluctuating dust emission  \citep{ 10.1051/0004-6361/201525733}, {suggesting} that the sublimation radius expands simultaneously with bright UV/optical and vice versa. 
{It is crucial to note} that there {may be deviations} from this simple scenario suggesting that the inner dust torus did not reach an equilibrium state immediately following the UV/optical flux change  \citep{10.1088/0004-637X/788/2/159}. 

{For the sake of simplicity and generality}, we assume that luminosity of {an} AGN, as {a} sinusoidally pulsating source of emission,  is $L=\overline{L}(1+\sin \frac{2\pi t}{P})$ with average luminosity $\overline{L}$ \citep{10.1093/mnras/stx1269}, and that $\Rsub$ {is a dimension of dust ring.}
{The} dust time lag $\tau_{sub}$ {then} scales with luminosity L as
$\Rsub\propto\tau_{sub}/c\propto L^{0.5}$\citep{10.1088/2041-8205/784/1/L4}. A simple algebraic manipulation of previous equations {results in} the formulation of dust radius fluctuation as follows:

\begin{equation}
\mathbf{r}_{\rm sub}=\overline{\mathbf{r}}_{\rm sub}\sqrt{\left(1+A\sin \frac{2\pi t}{P}\right)}
,\end{equation}
\noindent {where} the average dust ring offset is $\overline{\mathbf{r}}_{\rm sub}$, the amplitude is $A,$ and the period of the orbital motion of \eCBSMBH\, is $P$. We {suppose} that the amplitude scales as $A\propto \frac{P}{2\pi \tau_{d}}\sin{\frac{2\pi \tau_d}{P}}$ where $\tau_{d}=\frac{\Rsub}{c}$, {as with a} sinusoidally pulsating source of emission \citep{10.1093/mnras/stx1269}.

\end{appendix}
\end{document}